\documentclass[conference]{IEEEtran}
\IEEEoverridecommandlockouts
\usepackage{cite}
\usepackage{amsmath,amssymb,amsfonts}
\usepackage{algorithmic}
\usepackage{graphicx}
\usepackage{textcomp}
\usepackage{xcolor}
\usepackage[linesnumbered,ruled,vlined]{algorithm2e}
\usepackage{booktabs} 
\usepackage{subcaption}
\usepackage[table]{xcolor}
\newtheorem{theorem}{\textbf{Theorem}}
\newenvironment{proof}{{\indent \indent \it Proof:\quad}}{\hfill $\blacksquare$\par}
\usepackage{lineno}     % 添加跳转链接
\usepackage{caption}

\usepackage{tikz}
\newcommand{\circlednum}[1]{%
	\tikz[baseline=(char.base)]\node[shape=circle,draw,inner sep=0.5pt] (char) {\footnotesize #1};%
}
\SetAlCapSkip{2pt}
\def\BibTeX{{\rm B\kern-.05em{\sc i\kern-.025em b}\kern-.08em
		T\kern-.1667em\lower.7ex\hbox{E}\kern-.125emX}}
\begin{document}
	\setlength{\columnsep}{0.23 in}
	
	\title{LLM-Enhanced Deep Reinforcement Learning for Task Offloading in Collaborative Edge Computing\\
		% \thanks{Identify the applicable funding agency here. If none, delete this.}
		% \thanks{\textsuperscript{1}The code is available at repository: \url{https://anonymous.4open.science/r/LeDRL-854B/}.}
	}

\author{
	\IEEEauthorblockN{
		Hao Guo$^{\dagger}$, 
		Kaixiang Xu$^{\dagger}$, 
		Ziwu Ge, 
		Lei Yang$^{*}$
	}
	\IEEEauthorblockA{School of Software Engineering, South China University of Technology, Guangzhou, China}
	\IEEEauthorblockA{Email: \{seguohao,202330551861,202330550381\}@mail.scut.edu.cn; $^{*}$sely@scut.edu.cn}
	\IEEEauthorblockA{$^{\dagger}$ These authors contributed equally to this work; $^{*}$ Corresponding author}
}
	
	\maketitle
	
	\begin{abstract}
		
		Collaborative edge computing uses edge nodes in different locations to execute tasks, necessitating dynamic task offloading decisions to maintain low latency and high reliability, especially under unpredictable node failures. Although deep reinforcement learning (DRL) and large language models (LLMs) have shown promise for task offloading, DRL often suffers from poor sample efficiency and local optima, while LLMs are difficult to use directly due to inference overhead and output uncertainty.
		To address these limitations, we propose \textbf{LeDRL}, a hybrid decision framework that couples a \emph{lightweight LLM} with self-attention-enhanced DRL for real-time task offloading. LeDRL constructs structured, context-aware prompts capturing node status, task semantics, and link dynamics to derive high-level strategy priors. These are selectively processed by a self-attention-based alignment module for context-aware policy optimization. A reflective evaluator further distills semantic feedback from past trajectories to refine subsequent prompts and provide consistent guidance.
		Extensive experiments show that LeDRL outperforms representative baselines in task success rate, convergence speed, and real-time responsiveness across diverse network scales, achieving over 17\% improvement in success rate. Furthermore, we deploy LeDRL on Jetson-based edge devices using our prototype system \textit{CoEdgeSys}, demonstrating its robustness and feasibility under resource constraints.
		Our code is available at:https://github.com/GalleyG5/LeDRL.git.

	\end{abstract}
	

	\section{Introduction}
	Edge nodes distributed across different locations can cooperate to execute tasks, improving resource utilization and users' Quality of Experience (QoE), a paradigm known as Collaborative Edge Computing (CEC). A fundamental challenge in this context is task offloading: deciding whether to execute locally or forward to neighboring nodes to minimize latency while ensuring reliability~\cite{lin2024decentralized}. Many existing solutions rely on static heuristics or pre-trained policies that lack adaptability to runtime dynamics. Consequently, such methods are brittle in real-world environments where node failures, link disruptions, and topology changes are frequent~\cite{chinaev2023online}, leading to task failures and degraded performance.

\begin{figure}
		\centerline{
			\includegraphics[clip, width=0.85\linewidth]{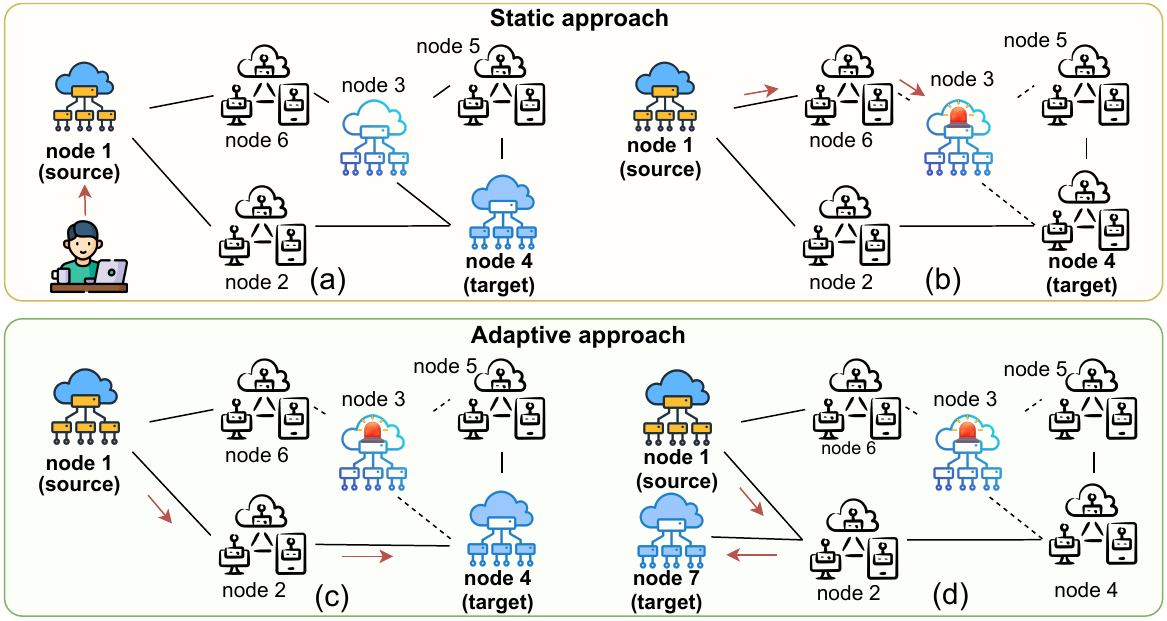}
		}
		\captionsetup{font={scriptsize,stretch=1.25}, justification=raggedright, singlelinecheck=false}
		\caption{
			\textbf{Adaptive vs. static task offloading in a detection scenario.} (a) Node 1 offloads to Node 4. (b) At $t_1$, Node 3 fails—static strategy cannot adapt. (c) Adaptive method reroutes tasks via alternate paths based on runtime conditions. (d) At $t_2$, Node 7 joins; the adaptive strategy leverages its low load and proximity to improve latency.
		}
		\label{fig:introduction}
		%	\vspace{-10pt}
	\end{figure}

	In contrast, adaptive offloading algorithms monitor node and network states to dynamically adjust decisions. As shown in Fig.~\ref{fig:introduction}, we illustrate static vs. adaptive offloading in a YOLO-based detection scenario. Initially, Node 1 attempts to offload a task to Node 4 (Fig.~\ref{fig:introduction}(a)). At time $t_1$, a failure at Node 3 causes transmission failure and inference delay under static policies (Fig.~\ref{fig:introduction}(b)). Adaptive methods, by contrast, can infer potential risks and reroute tasks through more stable paths (Fig.~\ref{fig:introduction}(c)). Later, at $t_2$, Node 7 with lower load and closer proximity joins the network, and the adaptive strategy redirects the task to reduce latency (Fig.~\ref{fig:introduction}(d)). These dynamics underscore the need for intelligent decision-making responsive to runtime uncertainties in edge environments.
	
	Deep Reinforcement Learning (DRL) has demonstrated strong potential for autonomous task offloading in CEC settings~\cite{kuru2023definition,lin2024decentralized}. Its strength lies in learning adaptive policies through continuous interaction with dynamic environments. However, DRL still faces key limitations. First, high sample complexity leads to inefficient training. 
	Second, in large-scale networks, the state and action spaces grow rapidly due to the curse of dimensionality, which slows convergence~\cite{kang2024imitation} and hurts QoE for latency-sensitive tasks.
	Moreover, node failures and topology shifts change the set of valid actions, often resulting in sub-optimal or invalid decisions. These issues limit the scalability and robustness of DRL in real-world edge deployments.

	Recently, Large Language Models (LLMs) have shown impressive capabilities in contextual reasoning and knowledge generalization~\cite{yang2025qwen3}. Emerging studies suggest that the prior knowledge encoded in LLMs can offer valuable heuristic guidance for task offloading~\cite{zhu2024task,song2024task}. Nonetheless, using LLMs as standalone decision-makers faces two critical challenges: (i) the substantial inference latency introduced by their extensive parameterization often violates the stringent real-time requirements of edge computing, and (ii) their stochastic output behaviors can lead to inconsistent and unreliable decisions. These limitations hinder their direct applicability in delay-sensitive and resource-constrained scenarios.
	%~\cite{cui2024survey}

	To tackle these challenges, we present \textbf{LeDRL}, a hybrid decision framework that unifies LLM-guided semantic abstraction and DRL policy adaptation through an attention-driven policy fusion mechanism that aligns LLM-derived priors with local observations.
	Unlike existing studies, LeDRL introduces a \emph{Reflective Evaluator} that transforms episodic task outcomes into structured semantic feedback, guiding policy refinement through contextual memory. A \emph{self-attention-based fusion module} is further designed to align LLM-generated intents with local observations, enabling the DRL agent to selectively absorb semantic knowledge and adapt to non-stationary environments.
	% 第二点
	To balance semantic guidance and real-time performance, LeDRL integrates a \textbf{lightweight LLM} into the online decision loop. The compact LLM provides high-level cues to enhance DRL adaptability, while keeping inference latency low. This design allows LeDRL to retain the reasoning benefits of LLMs while maintaining the speed and scalability required for real-time edge systems.
	The main contributions are summarized as follows:
	
	\begin{itemize}
		\item We model task offloading in collaborative edge environments as a decentralized partially observable Markov decision process (Dec-POMDP), and formulate it as a Mixed Integer Nonlinear Programming (MINLP) problem, which is proven to be NP-hard.
		
		\item We propose \textbf{LeDRL}, a novel hybrid framework that integrates a \textbf{lightweight language model} with DRL-based policy learning, enabling real-time, language-guided decision-making. It incorporates a self-attention-based policy alignment module to fuse LLM priors with local observations, enabling sample-efficient and robust adaptation to edge dynamics.
		
		\item We design a context-aware reflective evaluator that converts episodic task outcomes into structured semantic feedback to guide DRL policy refinement.
		
		\item Extensive evaluations show that LeDRL outperforms DRL and LLM-based baselines in success rate and convergence speed. We further implement a prototype system and deploy LeDRL on Jetson-based edge devices, demonstrating real-world feasibility and scalability.
	\end{itemize}

	\section{RELATED WORK}
	
	DRL has been extensively applied to edge task offloading by formulating it as an MDP~\cite{dong2024task}, enabling adaptive decision-making under dynamic workloads. Prior works, such as DAG-aware optimization~\cite{wang2021dependent} and log-driven cold-start training~\cite{lin2024decentralized}, focus on latency-energy trade-offs and stability. However, these methods assume static topologies, limiting their applicability in real-world edge systems with dynamic links and nodes.
	To improve robustness, task offloading has explored meta-heuristics for UAVs~\cite{nguyen2023dependency}, and mobility-aware routing~\cite{zhao2023meson}. Yet, most operate under fixed infrastructure assumptions and lack adaptability to heterogeneous, time-varying edge environments.
	
	Recently, LLMs have been introduced to edge intelligence for their semantic reasoning and generalization capabilities. Song et al.~\cite{song2024task} guided single-hop offloading using prompt-based LLMs, while Tong et al.~\cite{tong2025wirelessagent} proposed WirelessAgent, an LLM-powered agent integrating memory and planning for wireless optimization. Liu et al.~\cite{liu2024resource} focused on stabilizing edge inference by co-optimizing communication and computation. 
	However, these works mainly address system-level orchestration or static inference, without modeling decentralized multi-hop offloading under dynamic topologies.

	A few recent studies begin to explore LLM-DRL hybrid paradigms. For instance, Zhu et al.~\cite{zhu2024task} introduced a QTRAN–LLM framework with attention-based graph modeling for UAV task cooperation, and Ren et al.~\cite{ren2024industrial} used LLM-derived cognitive signals to replace handcrafted rewards in IIoT reinforcement learning. These approaches show the promise of combining LLM expressiveness with DRL adaptability, but still rely on centralized training or stable settings.
	
	In contrast, we propose \textbf{LeDRL}, a decentralized task offloading framework for dynamic edge networks. We model offloading as a Dec POMDP, where each edge node is an agent that acts under partial observations. LeDRL brings LLM semantic guidance into DRL training as decision priors to speed up learning and improve adaptation. We also design a topology aware masking mechanism to keep the observation input consistent when nodes or links change. This design supports robust and scalable coordination under node failures, arrivals, and volatile links.

	\section{System Model and Problem Formulation} \label{sec:system_model}
	
	\subsection{System Model}
	
	\begin{figure}[htbp]
		\centering
		\includegraphics[width=0.95\linewidth]{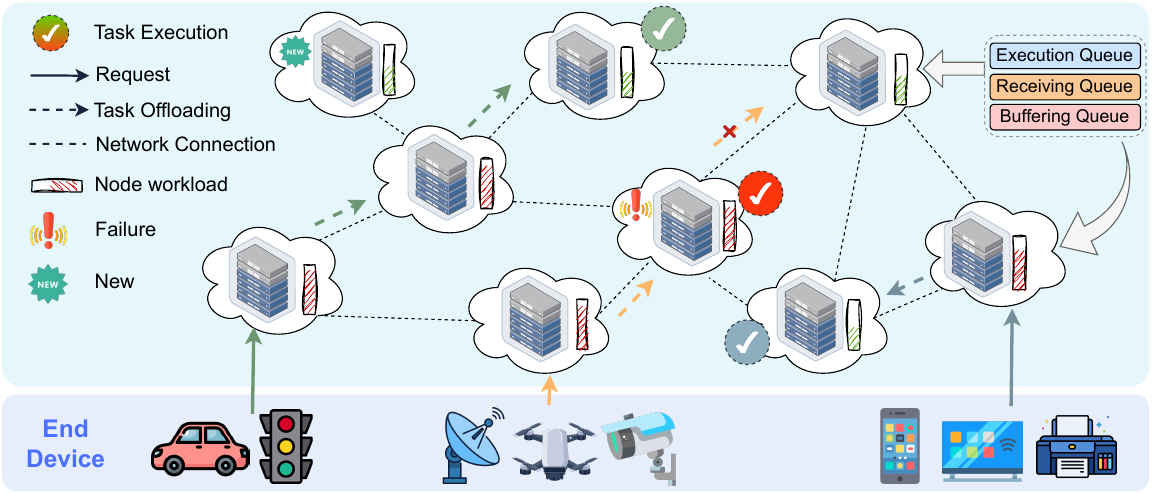}
		\captionsetup{font={scriptsize,stretch=1.25}, justification=raggedright, singlelinecheck=false}
		\caption{\textbf{System overview.} Tasks arrive at distributed edge nodes. Each node maintains local execution and communication queues, and a task can be processed locally or forwarded over multiple hops before execution at a destination node.}
		\label{fig:system-model}
		\vspace{-6pt}
	\end{figure}
	
	We consider a collaborative edge system modeled as a time-varying undirected graph $\mathcal{G}(t)=(\mathcal{V}(t),\mathcal{E}(t))$, where $\mathcal{V}(t)$ and $\mathcal{E}(t)$ denote the active nodes and available bidirectional links at slot $t$, respectively. Each node $v_i\in\mathcal{V}(t)$ has computing capacity $F_{v_i}$ and maintains $m_i$ communication queues, one for each of its $m_i$ neighboring nodes; the transmission rate on link $(v_i,v_j)$ is $R_{v_i v_j}$. As shown in Fig.~\ref{fig:system-model}, the basic offloading unit is an indivisible and independent task~\cite{lin2024decentralized}, which matches common edge workloads such as sensing records and image frames. More complex workflows, such as DAG applications, are assumed to be pre-partitioned into such atomic tasks before online scheduling.
	
	Time is slotted with equal duration and indexed by $t\in\{1,2,\ldots,T\}$. In each slot, node $v_i$ receives a task $K_i(t)$ with probability $\lambda_i$, following a Bernoulli arrival process~\cite{tang2020deep}. For task $K_i(t)$, let $S_i(t)$ denote its input size and $\Delta_i(t)$ its computation intensity in CPU cycles per bit~\cite{guo2025hybrid}$;$ the required CPU cycles are $C_i(t)=\Delta_i(t)\cdot S_i(t).$
	If the task generated at node $v_i$ is executed at node $v_j$, its execution delay and one-hop transmission delay are
	\begin{equation}
		T_{ij}^c(t)=\frac{C_i(t)}{F_{v_j}}, \qquad
		T_{ij}^t(t)=\frac{S_i(t)}{R_{v_i v_j}}.
	\end{equation}
	Besides service time, the task may wait in the transmission queue from $v_i$ to $v_j$ or in the execution queue of node $v_j$. Let $T_{ij}^{tw}(t)$ and $T_{ij}^{cw}(t)$ denote these two waiting times. If the task is still queued in the next slot, the corresponding waiting time increases by one slot:
	\begin{equation}
		T_{ij}^{cw}(t+1)=T_{ij}^{cw}(t)+1,\qquad
		T_{ij}^{tw}(t+1)=T_{ij}^{tw}(t)+1.
		\label{eq:wait-time}
	\end{equation}
	Hence, the total waiting time $W_i$ consists of the transmission-side and computation-side queueing delays captured in~\eqref{eq:wait-time}.
	
	To characterize unreliability in dynamic edge environments, we consider software failures during task execution, hardware failures of the executing node, and transmission failures over wireless links. Following~\cite{guo2025hybrid}, all three are modeled as Poisson processes. Let $\alpha_j$, $\gamma_j$, and $\beta_{ij}$ denote the software failure rate of node $v_j$, the hardware failure rate of $v_j$, and the link failure rate between $v_i$ and $v_j$, respectively. Then, for a task executed at node $v_j$, the probability of no software failure during execution interval $[0,T_{ij}^c(t)]$ is
	\begin{equation}
		R_{ij}^c(t)=\mathbb{P}\!\left[N_j(T_{ij}^c(t))=0\right]=e^{-\alpha_j T_{ij}^c(t)},
	\end{equation}
	where $N_j(T_{ij}^c(t))$ is the number of software failures occurring during execution. Similarly, the probability that node $v_j$ remains hardware-operational over the same interval, and the probability that transmission from $v_i$ to $v_j$ succeeds during $T_{ij}^t(t)$, are
	\begin{equation}
		R_j^e(t)=e^{-\gamma_j T_{ij}^c(t)},\qquad
		R_{ij}^t(t)=e^{-\beta_{ij} T_{ij}^t(t)}.
	\end{equation}
	Since edge applications are delay-sensitive, retransmissions are bounded~\cite{1576544}; therefore, $\beta_{ij}$ is treated as an effective link failure rate that captures the residual transmission failure probability under a limited-retry mechanism.

	\subsection{Problem Formulation}
	\subsubsection{Decision Variable}
	Let $x_{ij}(t)\in\{0,1\}$ be a binary variable with $x_{ij}(t)=1$ if task $K_i(t)$ is executed at node $v_j$ and $x_{ij}(t)=0$ otherwise.

	\subsubsection{Delay Constraints}
	Each task may be processed by at most one node, that is, $\sum_{j\in\mathcal{V}} x_{ij}(t) \le 1$.
	So, the end-to-end delay must satisfy
	\begin{equation}
		\label{eq:delay2}
		\begin{split}
			T_i(t)
			= \sum_{j\in\mathcal{V}} x_{ij}(t)
			\big(
			T_{ij}^{cw}(t) + T_{ij}^{c}(t)
			+ T_{ij}^{tw}(t) + T_{ij}^{t}(t)
			\big) \\
			\le D_i(t).
		\end{split}
	\end{equation}
	
	\subsubsection{Reliability Constraints}
    Each task must satisfy its minimum reliability requirement $\Phi_i$:
	\begin{align}
		\label{eq:reliability}
		R_i(t)
		&= R_{ij}^{c}(t)\cdot R_{ij}^{t}(t)\cdot R_{j}^{e}(t) \notag \\
		&= \exp\!\Big(-(\alpha_j+\gamma_j)T_{ij}^{c}(t)-\beta_{ij}T_{ij}^{t}(t)\Big)
		\ge \Phi_i .
	\end{align}

	\subsubsection{Resource Constraints}
	The cumulative resource consumption must not exceed node capacity:
	\begin{equation}
		\label{eq:resources}
		\left\{
		\begin{aligned}
			&\sum_{t=1}^{T}\sum_{i=1}^{N} C_i(t)\, x_{ij}(t) \le F_{v_j}\,\mathcal{T},\\
			&\sum_{t=1}^{T}\sum_{i=1}^{N} S_i(t)\, x_{ij}(t) \le R_{v_i v_j}\,\mathcal{T},\\
			&\sum_{t=1}^{T}\sum_{i=1}^{N} S_i(t)\, x_{ij}(t) \le M_{v_j}.
		\end{aligned}
		\right.
	\end{equation}
	Here, \( \mathcal{T} \) is the duration of the time horizon, and \( M_{v_j} \) is the memory capacity of node \( v_j \).

	\subsubsection{Optimization Objective}
	Our objective is to maximize the task success rate over the duration $\mathcal{T}$:
	\begin{equation}
		\max_{\{x_{ij}(t)\}}
		\frac{\sum_{t=1}^{T}\sum_{i=1}^{N}\sum_{j=1}^{N}
			x_{ij}(t)\,\mathbb{I}\!\left(T_i(t)\!\le\!D_i(t),\,R_i(t)\!\ge\!\Phi_i\right)}
		{\sum_{t=1}^{T}\sum_{i=1}^{N}\sum_{j=1}^{N} x_{ij}(t)},
		\label{eq:objective}
	\end{equation}
	%subject to constraints \eqref{eq:delay2}--\eqref{eq:resources},
	where $\mathbb{I}(\cdot)$ is an indicator that equals $1$ if $T_i(t)\le D_i(t)$ and $R_i(t)\ge \Phi_i$, and equals $0$ otherwise.
	
	\subsubsection{Computational Complexity}
	We formulate task offloading as a mixed integer nonlinear program (MINLP)~\cite{zhou2022joint} by optimizing binary decisions $x_{ij}(t)$ under the delay, reliability, and resource constraints in \eqref{eq:delay2}--\eqref{eq:resources}. The indicator term with binary variables leads to a combinatorial search space, while the reliability constraint introduces nonlinearity.
	
	\begin{theorem}
		\textit{Problem \eqref{eq:objective} subject to \eqref{eq:delay2}--\eqref{eq:resources} is NP-hard.}
	\end{theorem}
	
	\begin{proof}
		We prove NP-hardness via a polynomial reduction from \emph{Bin Packing} (NP-complete).
		Given item sizes $\{a_i\}_{i=1}^{N}$, $m$ bins, and capacity $\mathcal{Q}$, consider a single-slot instance with $T=1$ and
		$\mathcal{V}=\{v_1,\ldots,v_m\}$.
		For each task, choose $S_i(1)$ and $\Delta_i(1)$ such that $C_i(1)=\Delta_i(1)S_i(1)=a_i$, and set $F_{v_j}\mathcal{T}=\mathcal{Q}$ for all $j\in\{1,\ldots,m\}$.
		Choose deadlines, reliability parameters, link rates, and memory capacities so that \eqref{eq:delay2}, \eqref{eq:reliability}, and the last two inequalities in
		\eqref{eq:resources} are inactive.
		Assume each generated task is assigned to exactly one node,
		\begin{equation}
			\label{eq:assign_eq}
			\sum_{j\in\mathcal{V}} x_{ij}(1)=1,\quad \forall i\in\{1,\ldots,N\},
		\end{equation}
		then \eqref{eq:resources} reduces to, for each $j\in\{1,\ldots,m\}$,
		\begin{equation}
			\label{eq:bp}
			\sum_{i=1}^{N} a_i x_{ij}(1)\le \mathcal{Q}.
		\end{equation}
		With the mapping $x_{ij}(1)=1$ meaning that item $i$ is placed into bin $j$, feasibility is equivalent to Bin Packing.
		The reduction is polynomial, hence the problem is NP-hard.
	\end{proof}
	
%	\begin{theorem}
%		\textit{Problem \eqref{eq:objective} subject to \eqref{eq:delay2}--\eqref{eq:resources} is NP- hard.}
%	\end{theorem}
%
%	\begin{proof}
%		We prove NP-hardness via a polynomial reduction from \emph{Bin Packing} (NP-complete).
%		Given item sizes $\{a_i\}_{i=1}^{N}$, $m$ bins, and capacity $\mathcal{Q}$, consider a single-slot instance with $T=1$ and
%		$\mathcal{V}=\{v_1,\ldots,v_m\}$.
%		Set $C_i(1)=a_i$ for all $i\in\{1,\ldots,N\}$ and $F_{v_j}\mathcal{T}=\mathcal{Q}$ for all $j\in\{1,\ldots,m\}$.
%		Choose parameters so that \eqref{eq:delay2} and \eqref{eq:reliability} always hold, and make the last two inequalities in
%		\eqref{eq:resources} inactive by setting $S_i(1)=0$ and taking $M_{v_j}$ large enough.
%		Assume each generated task is assigned to exactly one node,
%		\begin{equation}
%			\label{eq:assign_eq}
%			\sum_{j\in\mathcal{V}} x_{ij}(1)=1,\quad \forall i\in\{1,\ldots,N\},
%		\end{equation}
%		then \eqref{eq:resources} reduces to, for each $j\in\{1,\ldots,m\}$,
%		\begin{equation}
%			\label{eq:bp}
%			\sum_{i=1}^{N} a_i x_{ij}(1)\le \mathcal{Q}.
%		\end{equation}
%		With the mapping $x_{ij}(1)=1$ meaning that item $i$ is placed into bin $j$, feasibility is equivalent to Bin Packing.
%		The reduction is polynomial, hence the problem is NP-hard.
%	\end{proof}
%	

	\begin{figure*}
		\centering
		\includegraphics[clip, width=0.9\textwidth]{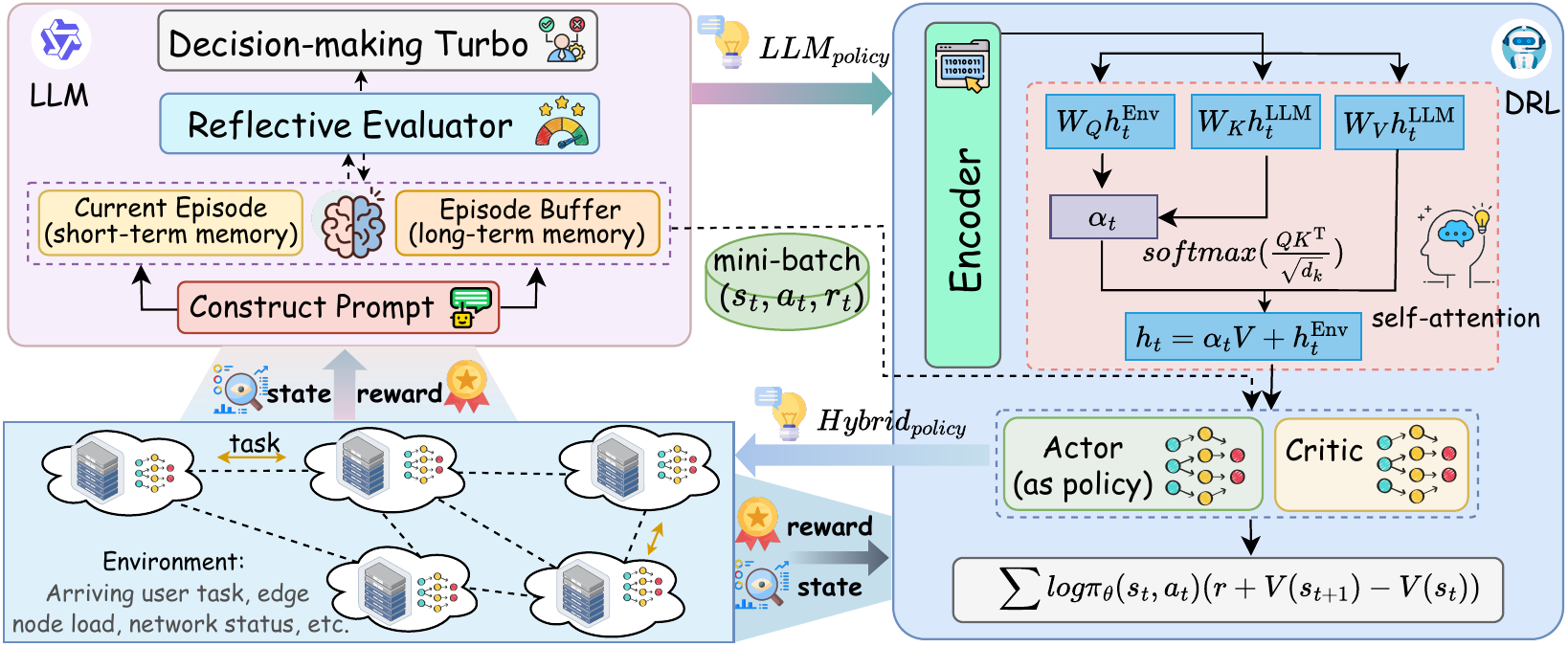}
		\captionsetup{font={scriptsize,stretch=1.25}, justification=raggedright, singlelinecheck=false}
		\caption{\textbf{Overview of the LeDRL framework}. An LLM provides semantic guidance during DRL decision-making. A self-attention fusion module merges LLM guidance with the DRL policy, and the RL agent outputs a hybrid offloading decision.}
		\label{fig:LEDRL}
		%	\vspace{-18pt}
	\end{figure*}

	\section{APPROACH DESIGN} \label{sec:approach}
	
%	LeDRL combines LLM priors with DRL exploration to improve decision quality, sample efficiency, and robustness in dynamic edge environments. As shown in Fig.~\ref{fig:LEDRL}, it contains two coupled components: an LLM-guided decision engine, including prompt construction, dual-memory retrieval, reflection, and a structured decision head; and a DRL agent, which aligns LLM guidance with local observations through an attention module and learns a stable decentralized policy under non-stationary conditions.
	
    In this section, we present LeDRL. Section~\ref{sec:Dec-POMDP} formulates the online collaborative edge offloading problem as a Dec-POMDP; Section~\ref{sec:LLM-guide} introduces an LLM-guided decision engine featuring prompt construction, dual-memory retrieval, reflection, and structured decision generation; and Section~\ref{sec:DRL-guided} presents a DRL agent that aligns LLM guidance with local observations through an attention module to learn a stable decentralized policy under non-stationary conditions.
	
	\subsection{Dec-POMDP Formulation} \label{sec:Dec-POMDP}
	
	To enable online offloading under stochastic arrivals and topology changes, we formulate the decision process as a Dec-POMDP. At time $t$, choosing action $a_i^t=j$ sets $x_{ij}(t)=1$ and $x_{ik}(t)=0$ for all $k\ne j$.
	
	\subsubsection{\textbf{State Observation}}
	
	At time $t$, agent $i\in\mathcal{N}$ receives a local observation $o_i^t\in\mathcal{O}_i$:
	\begin{equation*}
		\begin{aligned}
			o_i^t=\Big\{
			&[\lambda_i,\alpha_i,\gamma_i,F_{v_i},l_i^e(t),l_i^b(t)],\\
			&[S_i(t),C_i(t),D_i(t),H_i(t),W_i(t)],\;
			\{(R_{v_i v_j},\beta_{ij})\}_{j\in\mathcal{N}_i^t}
			\Big\},
		\end{aligned}
	\end{equation*}
	where $l_i^e(t)$ and $l_i^b(t)$ are the execution and buffer queue lengths, $H_i(t)$ is the current forwarding count, $W_i(t)$ is the accumulated waiting time, and $\mathcal{N}_i^t$ is the set of currently reachable neighbors. All features are normalized to $[0,1]$. If no task arrives, the task vector is set to $\mathbf{0}$.
	
	\subsubsection{\textbf{Action Space and Masking}}
	
	Each agent chooses $a_i^t\in\mathcal{A}_i^t$, where $a_i^t=i$ denotes local execution, $a_i^t=j\in\mathcal{N}_i^t$ denotes forwarding to neighbor $v_j$, and $a_i^t=\varnothing$ denotes no task arrival. To respect dynamic connectivity, the valid action set is
	\[
	\mathcal{A}_i^t=\{i,\varnothing\}\cup\{j\in\mathcal{N}\mid M_i^t(j)=1\},
	\]
	where $M_i^t(j)=1$ indicates that the link between nodes $i$ and $j$ is available at time $t$.
	
	\subsubsection{\textbf{State Transition}}
	
	The transition function $\mathcal{P}(s^{t+1}\mid s^t,\boldsymbol{a}^t)$ is driven by stochastic task arrivals, queue updates, node service rates, and evolving network connectivity. These dynamics jointly determine the realized delay and reliability, which are evaluated by Eq.~\eqref{eq:delay2} and Eq.~\eqref{eq:reliability}.
	
	\subsubsection{\textbf{Reward}}
	
	For each task $i$ at time $t$, we define the success indicator and violation indicator as
	\[
	s_i^t=\mathbb{I}\big(T_i(t)\le D_i(t),\ R_i(t)\ge \Phi_i\big),
	\]
	\[
	v_i^t=\mathbb{I}\big(T_i(t)> D_i(t)\ \text{or}\ R_i(t)< \Phi_i\big).
	\]
	The team reward at time $t$ is
	\begin{equation}
		\label{eq:rt_def}
		r_t=\sum_{i=1}^{N}(s_i^t-v_i^t).
	\end{equation}
	Under this design, maximizing the expected return is aligned with maximizing the objective in Eq.~\eqref{eq:objective}.
	
	\subsubsection{\textbf{Objective and Policy Factorization}} 
	
	Our goal is to learn a factorized joint policy
	\[
	\boldsymbol{\pi}_{\boldsymbol{\theta}}(\boldsymbol{a}^t\mid\boldsymbol{o}^t)
	=\prod_{i=1}^{N}\pi_{\theta_i}(a_i^t\mid o_i^t),
	\]
	which maximizes the expected cumulative reward:
	\begin{equation}
		\label{eq:dec_pomdp_obj}
		\begin{aligned}
			\boldsymbol{\pi}_{\boldsymbol{\theta}}^*
			=\arg\max_{\boldsymbol{\pi}_{\boldsymbol{\theta}}}\ &
			\mathbb{E}_{\boldsymbol{\pi}_{\boldsymbol{\theta}}}
			\left[\sum_{t=0}^{\infty}\gamma^t r_t\right] \\
			\text{s.t. }\ &
			a_i^t\sim\pi_{\theta_i}(\cdot\mid o_i^t),\quad \forall i\in\mathcal{N}.
		\end{aligned}
	\end{equation}
	
	\subsection{LLM for Context-Aware Decision Guidance} \label{sec:LLM-guide}
	
	\subsubsection{Prompt Construction}
	
	To activate domain-specific reasoning, we construct a structured prompt:
	\begin{equation}
		\mathcal{P}_t=[\mathcal{S}_t;\mathcal{T}_t;\mathcal{N}_t].
	\end{equation}
	Here, $\mathcal{S}_t$ encodes node status, including capacity $F_{v_i}$, queue levels, and failure rates $\alpha_i$ and $\gamma_i$; $\mathcal{T}_t$ includes task size $S_i(t)$, computation intensity $\Delta_i(t)$, deadline $D_i(t)$, forwarding count $H_i(t)$, and waiting time $W_i(t)$; and $\mathcal{N}_t$ captures link bandwidth $R_{v_i v_j}$ and failure rate $\beta_{ij}$.
	
	\subsubsection{Memory Repository}
	
	We maintain a dual-store memory $\mathcal{M}=\{\mathcal{M}_{\text{short}},\mathcal{M}_{\text{long}}\}$ to provide stable and reusable context. The short-term store records recent trajectories $\tau_t=\{(o_i^t,a_i^t,r_t)\}$, while the long-term store keeps compact summaries of typical failures and successful patterns. To control prompt length, recent trajectories are periodically summarized and moved from $\mathcal{M}_{\text{short}}$ to $\mathcal{M}_{\text{long}}$.
	%%%
	During prompt generation, a small context set is retrieved as
	\begin{equation}
		\mathcal{C}_t \leftarrow \text{Retrieve}(o_i^t,\mathcal{M}), \quad
		\mathcal{P}_t \gets [\mathcal{S}_t;\mathcal{T}_t;\mathcal{N}_t;\mathcal{C}_t].
	\end{equation}
	A memory item $m$ is ranked by
	\begin{align}
		\text{Rel}(o_i^t,m)
		&=
		w_1\,\text{Loc}(v_i,m)+
		w_2\,\text{Type}(K_i(t),m)\nonumber\\
		&+
		w_3\,\text{Sim}_{\text{task}}(\mathcal{T}_t,m)+
		w_4\,\text{Sim}_{\text{load}}(\mathcal{S}_t,m),
	\end{align}
	where $\text{Loc}$ enforces node-local retrieval, $\text{Type}$ matches task type, and $\text{Sim}_{\text{task}}$ and $\text{Sim}_{\text{load}}$ favor similar task profiles and queue conditions. The top-$K$ items are returned as $\mathcal{C}_t$.
	
	\subsubsection{Reflective Evaluator}
	
	To reduce risky decisions with limited overhead, we build a short risk summary $\mathcal{R}_t$ from observable signals. For local execution, $\mathcal{R}_t$ includes expected waiting time and execution failure risk under current load. For offloading, it includes estimated transmission delay, link failure risk, and target-side load. If the outcome is poor, e.g., $r_t<0$, we trigger LLM introspection with
	\begin{equation}
		\texttt{Diag}_t=\{o_i^t,\ a_i^t,\ r_t,\ \mathcal{R}_t\},
	\end{equation}
	which asks for the likely cause of the failure or violation and a safer choice for similar states. The returned reflection is appended to $\mathcal{M}_{\text{long}}$ through a FIFO queue.
	
	\subsubsection{Decision-Making Turbo}
	
	The turbo module constructs a prompt at each decision step by integrating the current system observation, retrieved memory, and a risk summary. This prompt is then submitted to the LLM following a predefined output schema, which specifies that the response should indicate either local execution or offloading to a designated node. In cases where the LLM generates outputs that deviate from this schema, the system defaults to local execution to ensure safe operation.
	
%	At each step, the turbo module assembles a prompt from the current observation, retrieved memory, and risk summary, and queries the LLM with a fixed output schema:
%	\begin{quote}
%		\textit{You are a task scheduling expert in edge computing. Your output must follow one of the formats below:}\\
%		- Execute locally\\
%		- Offload to node \{ID\}
%	\end{quote}
%	Malformed outputs are defaulted to local execution for safety.
	
	\subsection{DRL Guided by LLM Policy} \label{sec:DRL-guided}
	
	To preserve DRL exploration while exploiting LLM guidance, we adopt a hybrid Actor--Critic framework based on MAPPO~\cite{yu2022surprising}, where the DRL policy $\pi_{\theta}$ is updated using both environmental feedback and LLM-informed priors.
	
	\subsubsection{Policy Optimization in DRL}
	
	For policy $\pi_{\theta}$, the state value, action value, and advantage functions are
	\begin{align}
		V^{\pi_{\theta}}(s_t) &= \mathbb{E}_{\pi_{\theta}}\!\left[\sum_{l=0}^{\infty}\gamma^l r_{t+l}\mid s_t\right], \\
		Q^{\pi_{\theta}}(s_t,a_t) &= \mathbb{E}_{\pi_{\theta}}\!\left[\sum_{l=0}^{\infty}\gamma^l r_{t+l}\mid s_t,a_t\right], \\
		A^{\pi_{\theta}}(s_t,a_t) &= Q^{\pi_{\theta}}(s_t,a_t)-V^{\pi_{\theta}}(s_t),
	\end{align}
	where $\gamma$ is the discount factor.
	
	The policy is optimized by the clipped surrogate objective
	\begin{equation}
		\mathcal{L}^{\text{CLIP}}(\theta)=\mathbb{E}_t\!\left[
		\min\big(\rho_t(\theta)\hat{A}_t,\ \text{clip}(\rho_t(\theta),1-\epsilon,1+\epsilon)\hat{A}_t\big)
		\right],
	\end{equation}
	where $\rho_t(\theta)=\frac{\pi_{\theta}(a_t\mid s_t)}{\pi_{\text{old}}(a_t\mid s_t)}$ is the probability ratio, $\hat{A}_t$ is the estimated advantage, and $\epsilon$ is the clipping parameter.
	
	We estimate $\hat{A}_t$ by generalized advantage estimation:
	\begin{align}
		\delta_t &= r_t+\gamma V_{\omega_{\text{old}}}(s_{t+1})-V_{\omega_{\text{old}}}(s_t), \\
		\hat{A}_t &= \sum_{l=0}^{n-1}(\gamma\lambda)^l\delta_{t+l}.
		\label{eq:gae}
	\end{align}
	Here, $\delta_t$ is the temporal-difference (TD) error, $V_{\omega_{\text{old}}}$ is the value function parameterized by $\omega_{\text{old}}$, and $\lambda$ controls the bias--variance trade-off.
	
	To encourage exploration, we add entropy regularization:
	\begin{equation}
		\mathcal{H}(\pi_{\theta})=-\sum_{a_t}\pi_{\theta}(a_t\mid s_t)\log\pi_{\theta}(a_t\mid s_t).
	\end{equation}
	The final DRL objective is
	\begin{equation}
		\label{eq:drl_obj}
		\mathcal{L}_{\pi_{\theta}}=-\mathcal{L}^{\text{CLIP}}(\theta)-\beta \mathcal{H}(\pi_{\theta}),
	\end{equation}
	where $\beta$ balances policy optimization and exploration.
	
	\subsubsection{LLM Policy Distillation via Attention}
	
	The raw LLM output is discrete and non-differentiable. To align it with DRL optimization, we introduce an attention module $A_{\theta_{\text{att}}}$ that distills the LLM policy and injects it into DRL learning.
	
	\paragraph{Feature Embedding}
	The environment observation $o_t$ is encoded by $f_{\theta_1}$, and the LLM decision $a_t^{\text{LLM}}$ is embedded by a BERT-based \cite{devlin2019bert} encoder $e(\cdot)$ followed by $f_{\theta_2}$:
	\begin{align}
		h_t^{\text{Env}} &= f_{\theta_1}(o_t), \\
		h_t^{\text{LLM}} &= f_{\theta_2}(e(a_t^{\text{LLM}})).
	\end{align}
	
	\paragraph{Attention Fusion}
	The query, key, and value vectors of the attention module are defined as:
	\begin{equation}
		Q=W_Q h_t^{\text{Env}}, \quad K=W_K h_t^{\text{LLM}}, \quad V=W_V h_t^{\text{LLM}}.
	\end{equation}
	The attention score is
	\begin{equation}
		\alpha_t=\text{Softmax}\left(\frac{QK^\top}{\sqrt{d_k}}\right),
	\end{equation}
	where $d_k$ is the key dimension. The fused feature is
	\begin{equation}
		h_t=\alpha_t V+h_t^{\text{Env}}
		=\text{Softmax}\left(\frac{QK^\top}{\sqrt{d_k}}\right)V+f_{\theta_1}(o_t).
	\end{equation}
	The distilled LLM feature is then
	\begin{equation}
		g_t^{\text{LLM}}=O_{\theta_3}(h_t).
	\end{equation}
	
	\paragraph{Hybrid Loss Optimization}
	The attention module is trained by
	\begin{equation}
		\mathcal{L}_{\text{feat}}=\frac{1}{B}\sum_{i=1}^{B}
		\left\|A_{\theta_{\text{att}}}(o_i,a_i^{\text{LLM}})-g_i^{\text{LLM}}\right\|_2^2,
	\end{equation}
	\begin{equation}
		\mathcal{L}_{\text{act}}=\frac{1}{B}\sum_{i=1}^{B}
		\left(-z_{\pi_{\theta}}^{i}+\log\sum_{j=1}^{K}e^{z_{\pi_{\theta}}^{i,j}}\right),
	\end{equation}
	\begin{equation}
		\mathcal{L}_{A_{\theta_{\text{att}}}}=\mathcal{L}_{\text{feat}}+w_c\mathcal{L}_{\text{act}},
		\label{eq:loss-attention}
	\end{equation}
	where $w_c$ balances the two loss terms, $B$ is the mini-batch size, $z_{\pi_{\theta}}^{i}$ is the logit of the selected action, and $\{z_{\pi_{\theta}}^{i,j}\}_{j=1}^{K}$ are the logits of $K$ candidate actions.
	
	\subsubsection{Fusion of LLM and RL Policies}
	
	To match the LLM-derived feature space, the DRL agent also projects its observation $o_t$ into a latent representation:
	\begin{equation}
		g_t^{\text{DRL}}=\phi_{\text{DRL}}(o_t),
		\label{eq:drl_proj}
	\end{equation}
	where $\phi_{\text{DRL}}$ is the DRL feature encoder. Then, the fused representation for policy execution is
	\begin{equation}
		g_t^{\text{F}}=(1-\lambda_{\text{llm}})g_t^{\text{DRL}}+\lambda_{\text{llm}}\cdot\phi_{\text{Last}}(g_t^{\text{LLM}}),
		\label{eq:g_t}
	\end{equation}
	where $\phi_{\text{Last}}(\cdot)$ projects LLM guidance into a DRL-compatible space. The fusion coefficient is updated as
	\begin{equation}
		\lambda_{\text{llm}}^{(t)}=
		\begin{cases}
			\min(\beta_{\lambda}\lambda_{\text{init}},\ \lambda_{\text{llm}}^{(t-1)}\eta), & \text{if } t \bmod T_{\lambda}=0, \\
			\max(\lambda_{\min},\ \lambda_{\text{llm}}^{(t-1)}\gamma_{\text{decay}}), & \text{otherwise},
		\end{cases}
	\end{equation}
	where $T_{\lambda}$ is the scheduled update interval, $\beta_{\lambda}$ caps the maximum fusion weight, and $\eta,\gamma_{\text{decay}}\in(0,1)$ control its decay. The final objective is
	\begin{equation}
		\min \left( \mathcal{L}_{\pi_\theta} + \mathcal{L}_{A_{\theta_{\text{att}}}} \right).
		\label{eq:total_loss}
	\end{equation}
	
	The pseudo-code of LeDRL is given in Algorithm~\ref{alg:ledrl}.
	
	\begin{algorithm}
		\caption{Policy Optimization in LLM-Enhanced DRL (LeDRL)}
		\label{alg:ledrl}
		\SetKwInOut{Input}{Input}
		\SetKwInOut{Output}{Output}
		
		\Input{
			Initialize DRL policy $\pi_\theta$, attention module $A_{\theta_{\text{att}}}$, memory $\mathcal{M}=\{\mathcal{M}_{\text{short}},\mathcal{M}_{\text{long}}\}$, fusion weight $\lambda_{\text{llm}}=\lambda_{\text{init}}$, LLM policy $\pi_{\text{LLM}}$, fusion update interval $T_{\lambda}$, total episodes $N$, and episode length $T$.
		}
		\Output{Optimized policy $\pi_\theta^*$.}
		
		\BlankLine
		\For{episode $\leftarrow 1$ \KwTo $N$}{
			Observe initial state $s_0$\;
			
			\For{$t\leftarrow 0$ \KwTo $T$}{
				Construct prompt $\mathcal{P}_t=[\mathcal{S}_t;\mathcal{T}_t;\mathcal{N}_t]$\;
				Retrieve context $\mathcal{C}_t\gets\text{Retrieve}(o_t,\mathcal{M})$\;
				Generate LLM decision $a_t^{\text{LLM}}\gets \pi_{\text{LLM}}(\mathcal{P}_t,\mathcal{C}_t)$\;
				
				Encode state: $h_t^{\text{Env}}\gets f_{\theta_1}(o_t)$\;
				Encode LLM response: $h_t^{\text{LLM}}\gets f_{\theta_2}(e(a_t^{\text{LLM}}))$\;
				Compute fused feature using Eq.~\eqref{eq:g_t}\;
				Sample action $a_t\sim\pi_\theta(g_t^{\text{F}})$\;
				
				Observe reward $r_t$ and next state $s_{t+1}$\;
				Store $(o_t,a_t,r_t)$ into $\mathcal{M}_{\text{short}}$\;
				\If{$r_t<0$}{
					Update long-term memory: $\mathcal{M}_{\text{long}}\gets \text{Reflect}(o_t,a_t,r_t,\mathcal{R}_t)$\;
				}
				
				\If{$t \bmod T_{\lambda}=0$}{
					$\lambda_{\text{llm}}\gets \min(\beta_{\lambda}\lambda_{\text{init}},\eta\lambda_{\text{llm}})$\;
				}
				\Else{
					$\lambda_{\text{llm}}\gets \max(\lambda_{\min},\gamma_{\text{decay}}\lambda_{\text{llm}})$\;
				}
				
				Compute advantage $\hat{A}_t$ using Eq.~\eqref{eq:gae}\;
				Update policy: $\theta\gets \theta-\nabla_{\theta}\mathcal{L}_{\pi_{\theta}}$\;
				Update attention: $\theta_{\text{att}}\gets \theta_{\text{att}}-\nabla_{\theta_{\text{att}}}\mathcal{L}_{A_{\theta_{\text{att}}}}$\;
			}
		}
	\end{algorithm}

	\section{EVALUATION} \label{sec:evaluation}
	
	\subsection{Experiment Setup}
	All experiments are conducted on a workstation with an AMD Ryzen Threadripper 3990X CPU, an NVIDIA RTX 4090 GPU, and 256 GB RAM.
	
	\textbf{Environment Configuration.} We simulate a decentralized edge system over 100 time slots~\cite{lin2024decentralized}, with \(10\!\sim\!20\) heterogeneous nodes, stochastic task arrivals, and task input sizes uniformly sampled from \([2000,4000]\) KB~\cite{lin2024decentralized}. Task complexity ranges from \([800,2400]\) cycles/bit, and deadlines are fixed at 4 s~\cite{guo2025hybrid}. The network topology is randomly generated, sparse, and connected to model partial inter-node connectivity. Each node has a 3 GHz CPU, and link rates are sampled from \([10,40]\) MB/s~\cite{guo2025hybrid}. To capture network dynamics, we set the node failure rate to 0.01 and the node appearance probability to 0.1~\cite{knight2011internet}.
	
	\textbf{Model Configuration.} LeDRL is implemented in \textit{PyTorch} and integrates \textbf{Qwen3-4B}~\cite{yang2025qwen3} to guide DRL policy learning. It adopts an actor--critic architecture with MLP-based networks of hidden size 64, and uses a self-attention module with embedding dimension 8, maximum length 512, and dropout 0.1. Following~\cite{guo2025hybrid}, PPO is trained with learning rate \(0.0004\), decay factor 0.99, and discount factor \(\gamma=0.99\), and is evaluated every 4 iterations.
	
	\subsection{Baseline Approaches}
	We compare LeDRL with six baselines, covering three DRL methods, two heuristics, and one LLM framework. 
	
	\textbf{VDN-TO}~\cite{sunehag2017value}: a cooperative MARL baseline that factorizes the joint Q-function into per-node utilities for local execution or neighbor forwarding.
	
	\textbf{MAPPO-TO}~\cite{yu2022surprising}: a policy-gradient baseline that trains node-local policies with clipped updates to improve robustness under non-stationarity.
	
	\textbf{MASAC-TO}~\cite{haarnoja2018soft}: a multi-agent SAC baseline adapted to discrete actions to match our offloading decision space.
	
	\textbf{RATC}. A lightweight heuristic~\cite{beraldi2020power} that samples two neighbors and selects one based on deadline and reliability.
	
	%\textbf{AGSP}~\cite{10606273}. A hybrid-PSO enginewhose adaptive genetic-simulated-annealing kernel we transplant to decide, upon task arrival, whether to execute locally or forward to the best neighbor.
	\textbf{AGSP}~\cite{10606273}. A hybrid-PSO engine. We transplant its adaptive genetic--simulated annealing kernel to our setting. Upon each task arrival, it decides whether to execute locally or forward the task to the best neighbor.

	\textbf{Reflexion}~\cite{shinn2023reflexion}. an LLM-driven baseline that prompts \textbf{Qwen3-4B} with structured task--state--outcome tuples and uses iterative self-feedback to refine decentralized offloading decisions during training.
	
	\begin{figure}
		% 第一幅子图（占50%宽度）
		\begin{subfigure}{0.49\linewidth}
			\centering
			\includegraphics[clip, width=\linewidth]{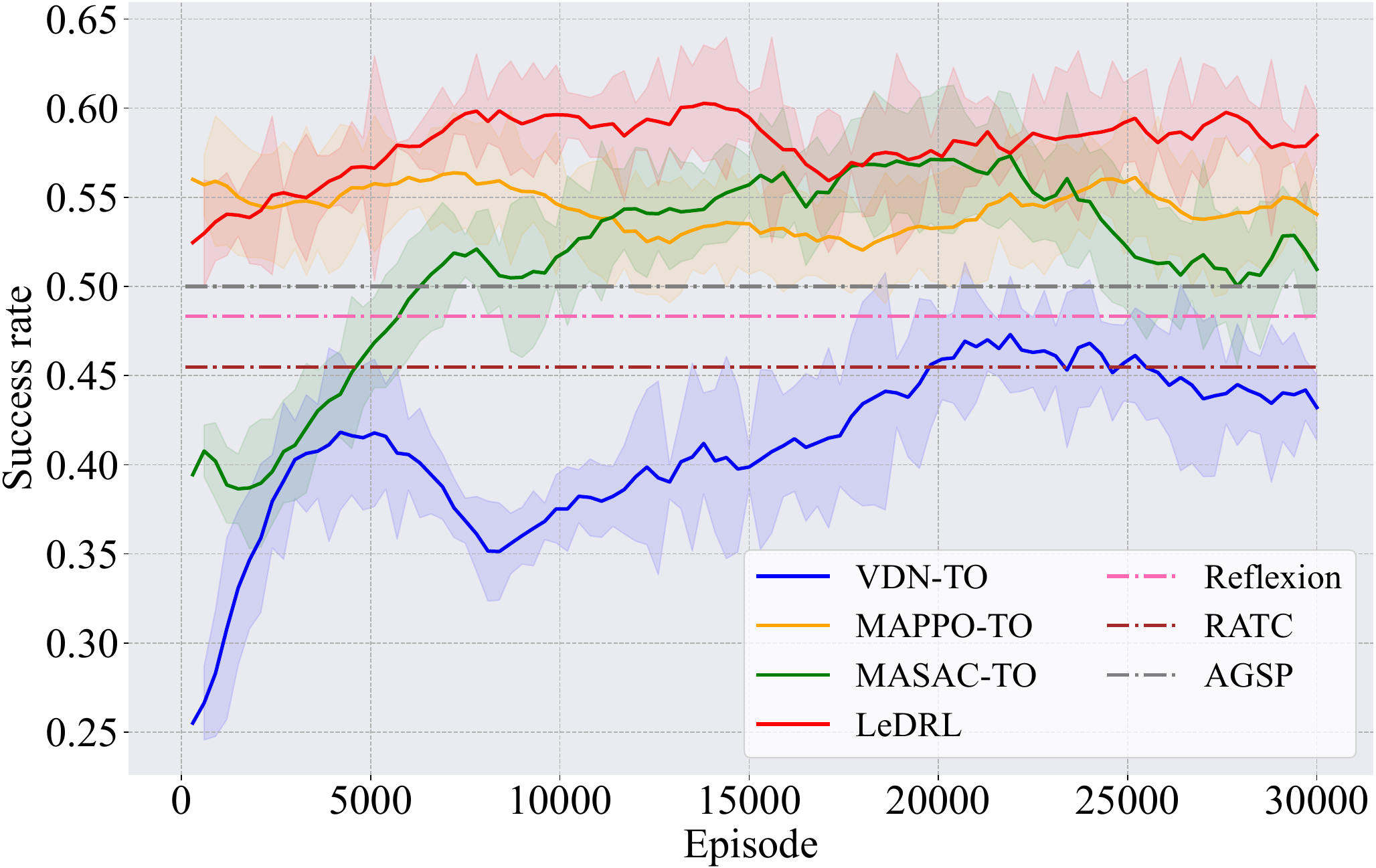}  % 宽度适应子图环境
			\captionsetup{font={scriptsize,stretch=1.25}}
			\caption{10-nodes}
			\label{fig:sub10}
		\end{subfigure}
		\hfill  % 两图之间留白
		% 第二幅子图（占50%宽度）
		\begin{subfigure}{0.49\linewidth}
			\centering
			\includegraphics[clip, width=\linewidth]{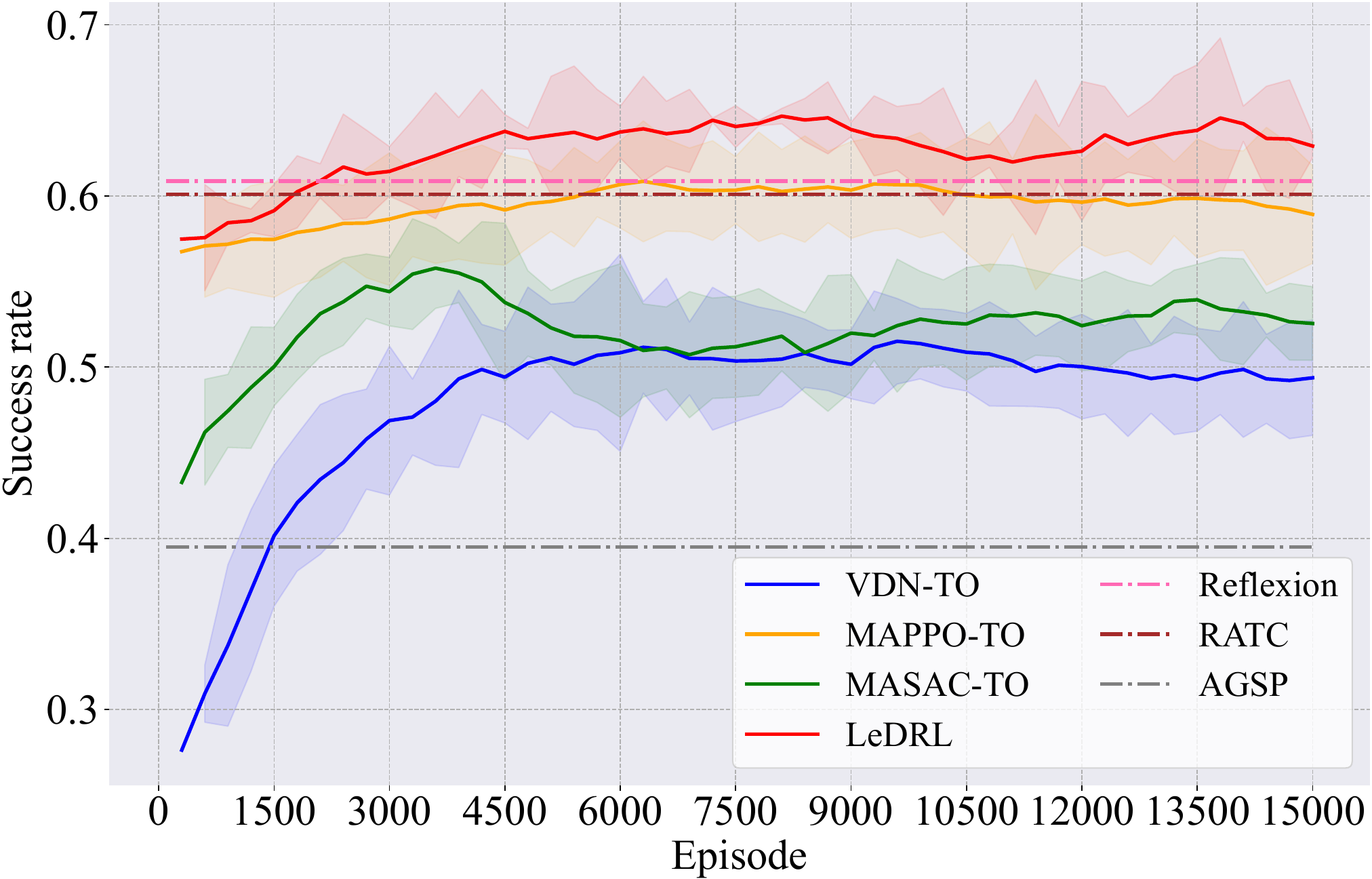}
			\captionsetup{font={scriptsize,stretch=1.25}}
			\caption{20-nodes}
			\label{fig:sub20}
		\end{subfigure}
		\vspace{-16pt}
		% 总标题
		\captionsetup{font={scriptsize,stretch=1.25}, justification=raggedright, singlelinecheck=false}
		\caption{Learning curves of task success rates under different methods for different network topologies.}
		\label{fig:node-comparison}
		%	\vspace{-16pt}
	\end{figure}
	
	%All baselines are carefully tuned and evaluated over 10 randomized trials per topology. We report mean task success rates with standard deviations to ensure fair and reproducible comparisons.
	All methods are tuned and evaluated on each topology over 10 randomized runs. We report mean task success rates with standard deviations for fair and reproducible comparison.

	\subsection{Performance on Simulation}

	% [htbp]
	\begin{figure*} 
		\captionsetup{aboveskip=-5pt} 
		\centering
		\begin{subfigure}{0.242\textwidth}
			\centering
			\includegraphics[width=1.7in,height=1.1in]{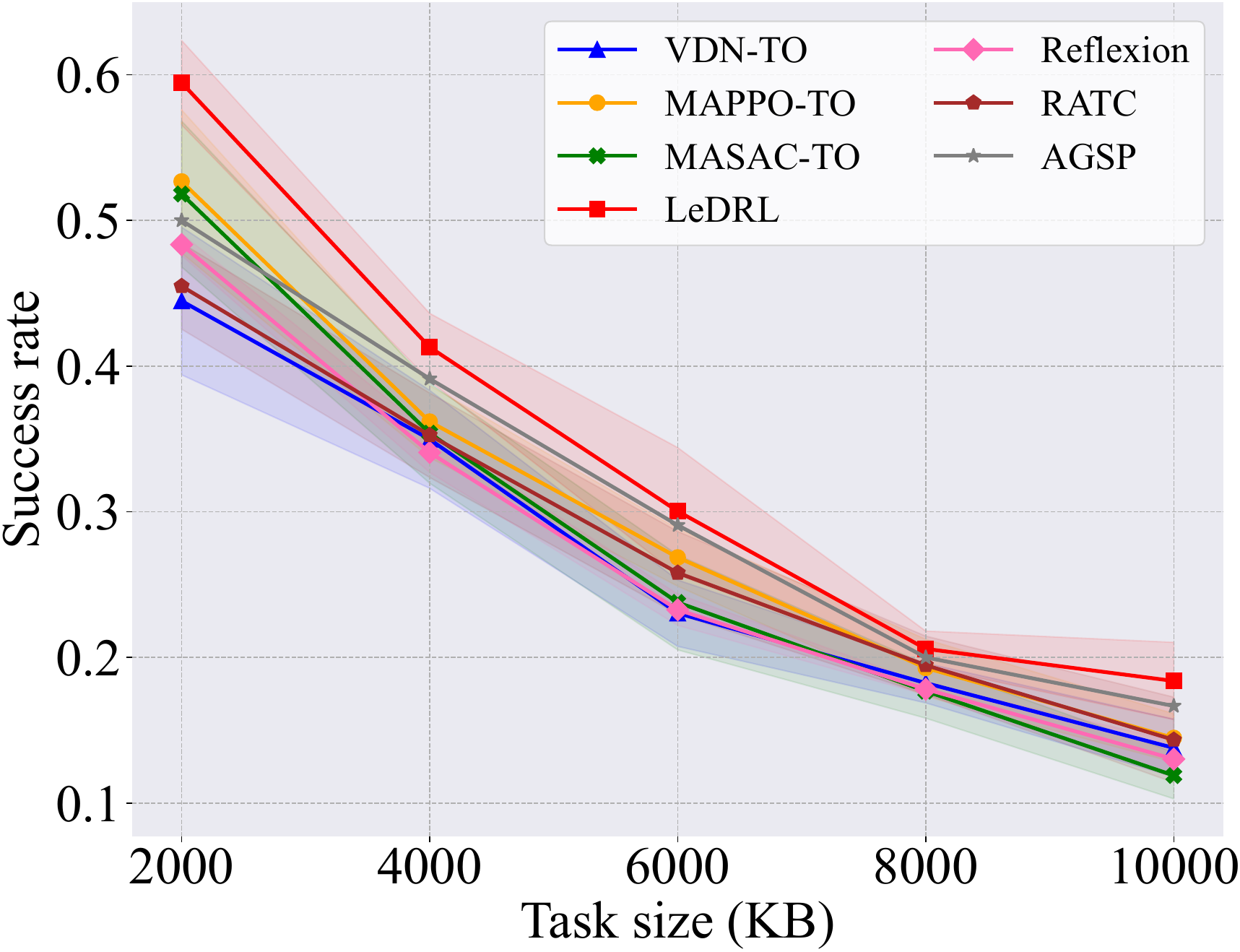}
			%		    	\caption{}
			\label{fig:robustness-task_size}
		\end{subfigure}
		\hfill
		\begin{subfigure}{0.242\textwidth}
			\centering
			\includegraphics[width=1.7in,height=1.1in]{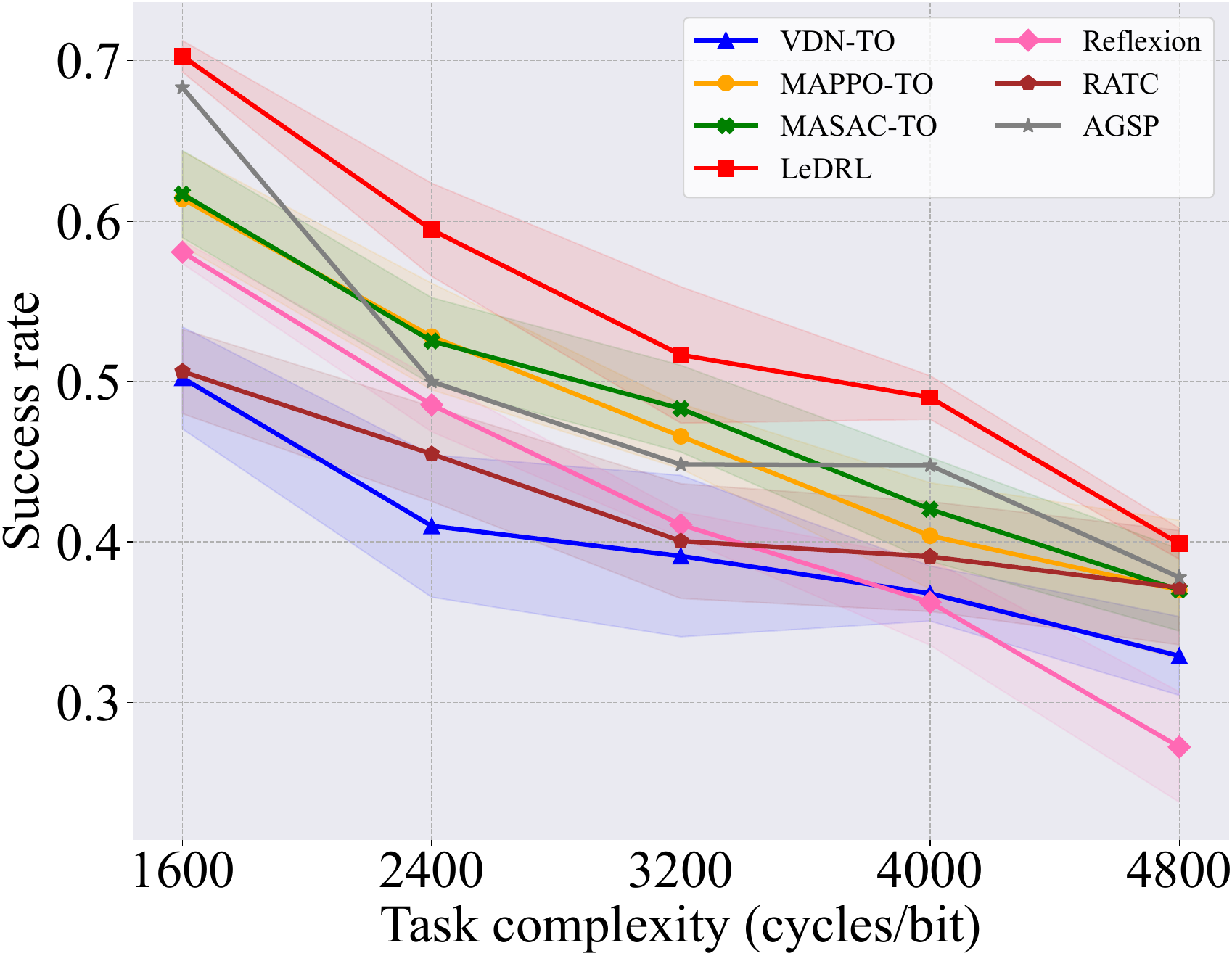}
			%		    	\caption{}
			\label{fig:robustness-task_complexity}
		\end{subfigure}
		\hfill
		\begin{subfigure}{0.242\textwidth}
			\centering
			\includegraphics[width=1.7in,height=1.1in]{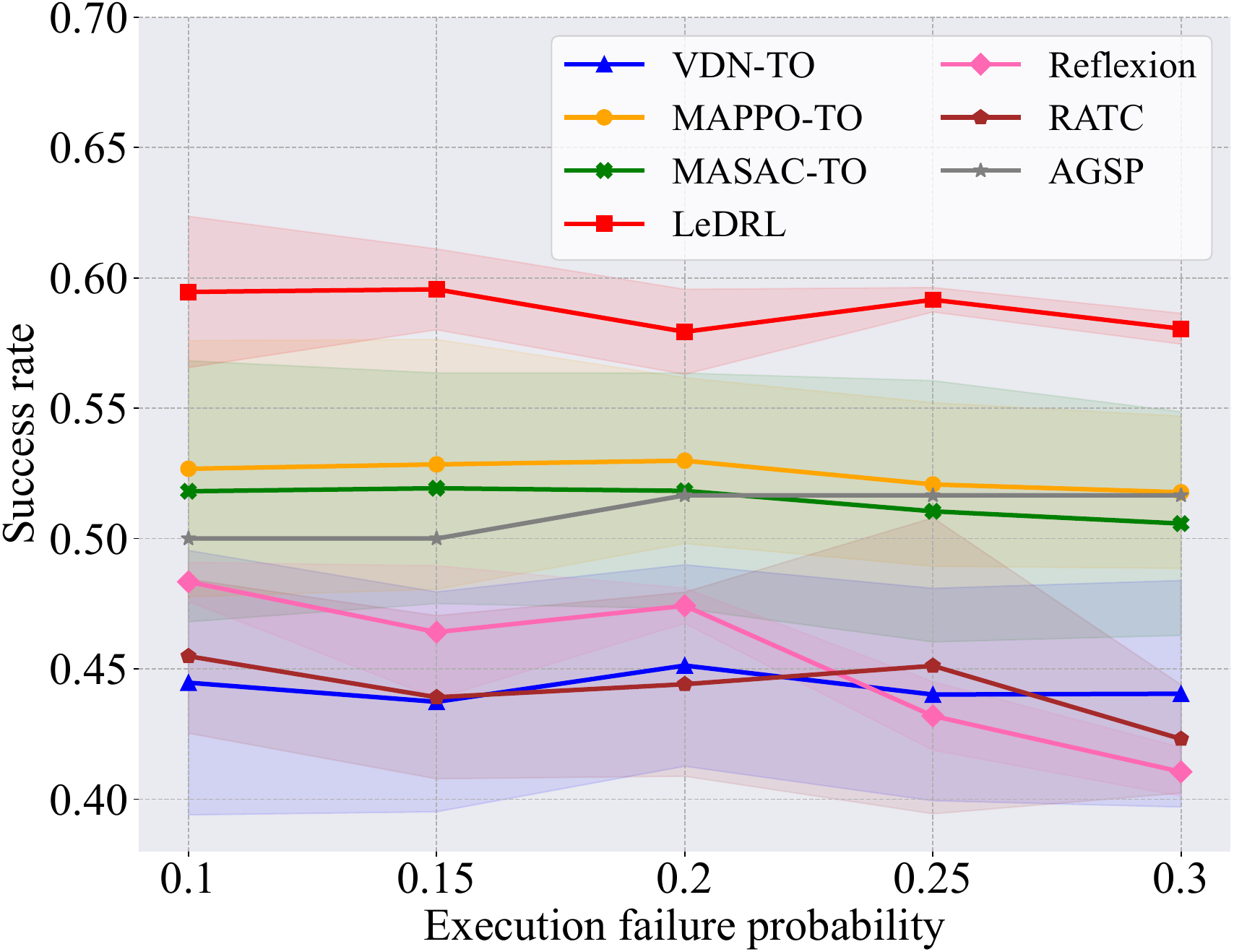}
			%		\caption{}
			\label{fig:robustness-exec}
		\end{subfigure}
		\hfill
		\begin{subfigure}{0.242\textwidth}
			\centering
			\includegraphics[width=1.7in,height=1.1in]{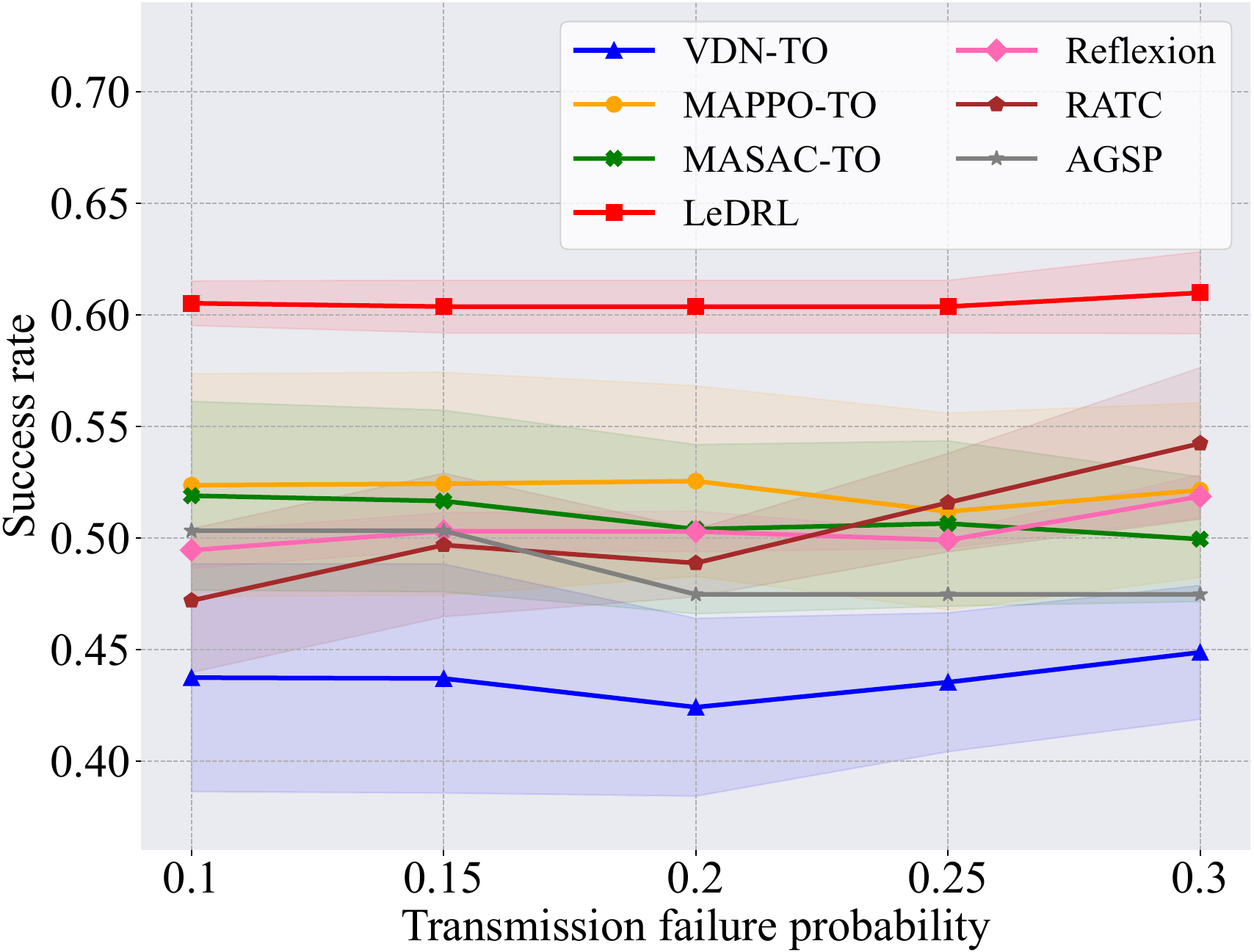}
			%		\caption{}
			\label{fig:robustness-trans}
		\end{subfigure}
		\hfill
		\captionsetup{font={scriptsize,stretch=1.25}, justification=raggedright, singlelinecheck=false}
		\caption{Success rate of tasks under different:(a) size; (b) complexity; (c) execution failure rates; (d) transmission failure rates.}
		\label{fig:robustness}
		\vspace{-16pt}
	\end{figure*}

	\subsubsection{Comparison of Training Performance} \label{sec:training-curves}
	
	Fig.~\ref{fig:node-comparison} shows the training curves of LeDRL and all baselines in terms of task success rate under different topologies. LeDRL converges faster, exhibits lower variance in the early stage of training, and achieves the highest final success rate in all settings. This directly supports our motivation: although standard DRL can be applied to task offloading, it often requires many samples and may become trapped in poor local optima in dynamic networks. For example, MAPPO-TO improves rapidly at the beginning but becomes unstable later and often struggles to escape sub-optimal regions without semantic priors. MASAC-TO learns quickly in small topologies but remains unstable, and in the 20-node case it often plateaus at a sub-optimal level, indicating limited scalability as the problem becomes more complex. VDN-TO is relatively stable but maintains a low success rate due to its limited adaptability.
	
	LeDRL improves sample efficiency by using LLM-guided priors to guide early exploration and enhance exploration quality. Reflective feedback further refines these priors, while the self-attention fusion module aligns semantic cues with local observations, thereby stabilizing learning under topology changes and node failures. By contrast, Reflexion, AGSP, and RATC can exploit structural information or handcrafted guidance, but they lack effective feedback-driven adaptation, so their performance varies more significantly and degrades in complex topologies.

	\subsubsection{Robustness Performance}
	
	We evaluate robustness on a fixed 10-node topology under four perturbations: task size, computation intensity, execution failures, and transmission failures (Fig.~\ref{fig:robustness}).
	As task input size and computation intensity increase, success rates generally decline because larger tasks and heavier computation increase transmission time, execution time, and queueing delay under limited bandwidth and computing resources, thereby causing more deadline violations, as shown in Fig.~\ref{fig:robustness}(a--b). LeDRL consistently performs best because LLM guidance provides feasible decision priors in the early stage, while the learned policy further adapts to local congestion and topology variations.
	
	For failure perturbations, several baselines show non-monotonic trends, with alternating rises and drops as failure probability increases, as shown in Fig.~\ref{fig:robustness}(c--d). This is expected in decentralized offloading because varying failure rates can shift the dominant bottleneck and change the relative benefit of forwarding versus local execution. Consequently, the same policy may perform differently across failure regimes, while stochastic arrivals and queueing dynamics further amplify fluctuations. Despite these shifts, LeDRL consistently achieves the highest success rate and better stability than the baselines. This advantage comes from reliability-aware LLM priors and self-attention fusion, which adaptively combine semantic guidance with local observations as failure patterns evolve. At a failure rate of 0.25, LeDRL improves success rate by about 12\% and 17\% over MAPPO-TO under execution-failure and transmission-failure settings, respectively. In contrast, VDN-TO and MASAC-TO are less reliability-aware and generalize poorly, Reflexion adapts slowly to fast disturbances, and RATC and AGSP rely on fixed rules and become unstable when the bottleneck shifts.

	\subsubsection{Impact of the network topology}
	%%%  单独占一行   [thbp]
	
	\begin{figure}
		% 第一幅子图（占50%宽度）
		\begin{subfigure}{0.49\linewidth}
			\centering
			\includegraphics[clip, width=\linewidth]{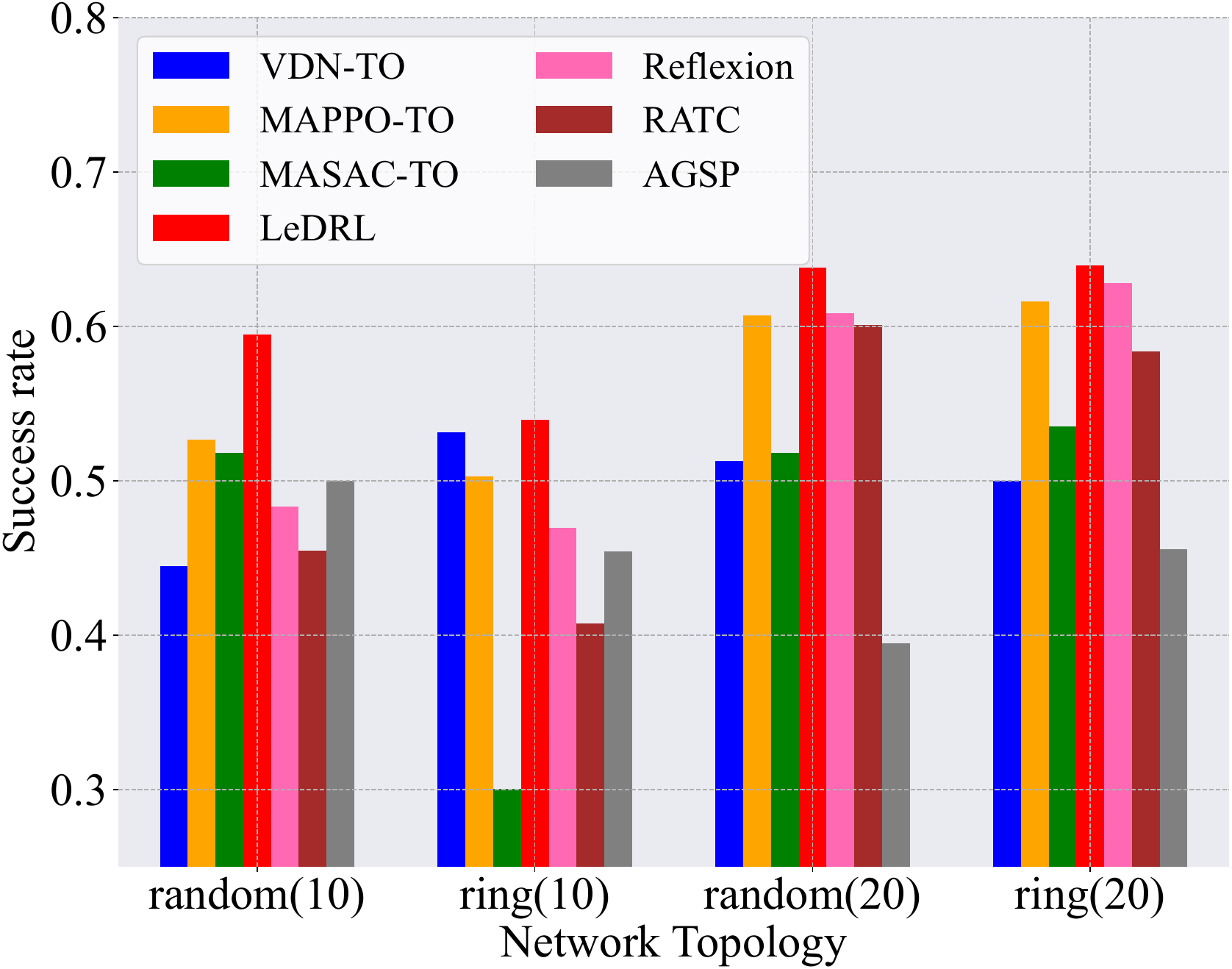}
			\captionsetup{font={scriptsize,stretch=1.25}}  
			\caption{Impact of the topology}
			\label{fig:topology}
		\end{subfigure}
		\hfill  % 两图之间留白
		% 第二幅子图（占50%宽度）
		\begin{subfigure}{0.49\linewidth}
			\centering
			\includegraphics[clip, width=\linewidth]{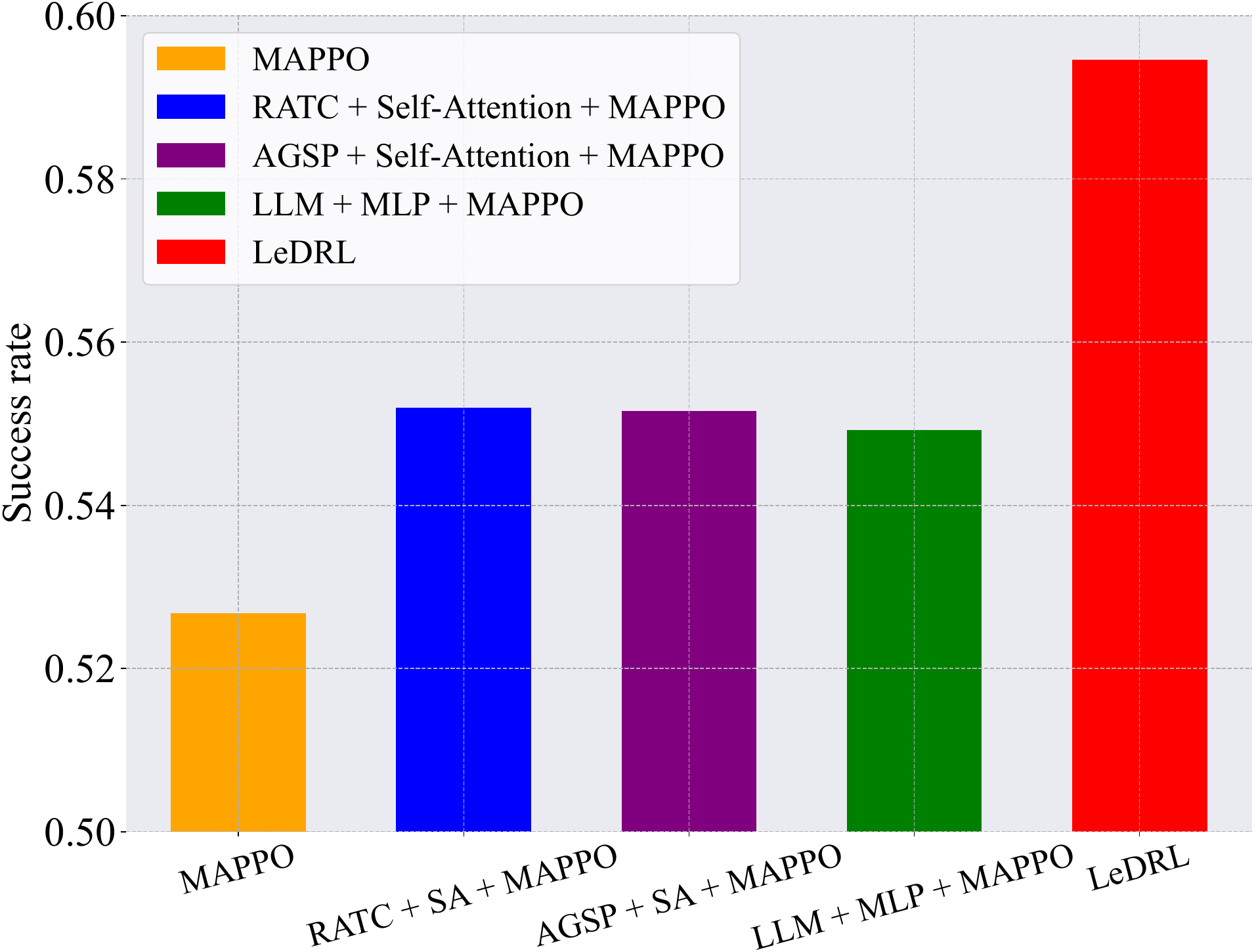}
			\captionsetup{font={scriptsize,stretch=1.25}}  
			\caption{Ablation study of LeDRL}
			\label{fig:methods_comparison}
		\end{subfigure}
		\vspace{-16pt}
		% 总标题
		\captionsetup{font={scriptsize,stretch=1.25}, justification=raggedright, singlelinecheck=false}
		\caption{Performance analysis of LeDRL: (a) network topology  and (b) ablation study of key components.}
		\label{fig:combined-analysis}
		%	\vspace{-16pt}
	\end{figure}

	\begin{table}[!t]
		\footnotesize 
		\centering
		\captionsetup[table]{aboveskip=3pt,belowskip=0pt}
		\caption{
			Task success rate and inference time across different models. 
			\textbf{Bold} indicates the best performance.
		}
		\vspace{-3pt}  % 减小标题与表格的间距
		\label{tab:results}
		\renewcommand{\arraystretch}{1.0}
		% 缩放表格至单栏宽度（\linewidth），高度自动适配
		\resizebox{\linewidth}{!}{
			\begin{tabular}{lcccc}
				\toprule
				& \multicolumn{2}{c}{10 Nodes} & \multicolumn{2}{c}{20 Nodes} \\
				\cmidrule(lr){2-3} \cmidrule(lr){4-5}
				\textbf{Model} & Success (\%) $\uparrow$ & Time (s) $\downarrow$ & Success (\%) $\uparrow$ & Time (s) $\downarrow$ \\
				\midrule
				RATC & 45.48 $\pm$ 1.24 & \textbf{0.0004} & 59.98 $\pm$ 1.45 & \textbf{0.0008} \\
				AGSP & 50.01 $\pm$ 1.63 & 0.0005 & 39.48 $\pm$ 1.72 & 0.0018 \\
				VDN-TO & 44.47 $\pm$ 1.92 & 0.0028 & 51.27 $\pm$ 1.82 & 0.0026 \\
				MASAC-TO & 51.81 $\pm$ 2.88 & 0.0025 & 51.80 $\pm$ 1.77 & 0.0024 \\
				MAPPO-TO & \textit{52.68 $\pm$ 3.40} & 0.0035 & 60.68 $\pm$ 2.24 & 0.0043 \\
				Reflexion & 48.34 $\pm$ 1.97 & 1.5786 & \textit{60.86} $\pm$ 1.39 & 3.0128 \\
				\rowcolor{gray!30}\textbf{LeDRL} & \textbf{59.46 $\pm$ 1.42} & 0.7046 & \textbf{63.78 $\pm$ 1.68} & 0.7379 \\
				\bottomrule
			\end{tabular}
		}
		%	\vspace{-10pt}
	\end{table}
	
	We evaluate LeDRL on two network scales (10 and 20 nodes) and two topology types: connected \textbf{random} graphs and structured \textbf{ring} topologies. As shown in Fig.~\ref{fig:topology}, LeDRL achieves the highest task success rate in all settings. In contrast, MAPPO-TO performs well in \textbf{random} graphs but drops markedly in \textbf{ring} topologies, where routing flexibility is limited and congestion can propagate along the ring. RATC and AGSP are even more sensitive to topology changes and show larger performance fluctuations, reflecting their reliance on fixed decision rules.
	
	LeDRL remains robust because it combines semantic guidance with adaptive learning. The LLM guidance provides topology-aware priors, such as avoiding risky relays and preferring feasible paths under tight delay and reliability constraints, which improves early exploration and reduces wasted interactions. The DRL policy then refines these priors using local observations, so it can react to queue build-up and link variability in different graphs. By abstracting task and network conditions into compact semantic cues, LeDRL reduces the effective decision complexity and improves generalization when scaling from 10 to 20 nodes and when switching between random and ring structures.
	
	\subsubsection{Ablation Study}
	
	Fig.~\ref{fig:methods_comparison} compares LeDRL with variants that remove or replace key components. Plain MAPPO yields the lowest success rate, as it relies on local trial-and-error and has no explicit mechanism to handle long-horizon risk under dynamic edge conditions. Adding LLM guidance with a simple MLP (e.g., LLM+MLP+MAPPO) improves performance, but the gain is limited because raw semantic hints are not reliably aligned with the policy input. Replacing the MLP with self-attention (e.g., RATC+SA+MAPPO and AGSP+SA+MAPPO) further improves success rate, since attention can better fuse heterogeneous signals and reduce policy drift. However, these variants still lack context-aware semantic refinement, so they converge to suboptimal policies.
	
	LeDRL achieves the best performance by closing this gap. Its structured prompts and context-aware memory retrieval provide relevant semantic priors, and the reflective evaluator updates these priors based on episodic outcomes. The self-attention fusion then selectively integrates the refined guidance with local observations, leading to faster convergence and more stable policies. 
	%Overall, the ablation results confirm that both semantic guidance and adaptive alignment are necessary to obtain consistent gains in edge offloading.

	\subsubsection{Inference Efficiency and Performance Trade-off} 
	
	Table~\ref{tab:results} reports task success rate and inference latency across different network scales. Heuristic methods such as RATC and AGSP achieve the lowest latency, but their success rates decline in dynamic settings because they cannot effectively adapt to failures or topology changes. LeDRL achieves the highest success rate in both 10- and 20-node networks while maintaining moderate inference latency, remaining compatible with the 4\,s delay requirement.
	
	This advantage comes from LeDRL’s collaborative inference design: a \textbf{lightweight LLM} provides semantic guidance at decision time, helping the policy select more reliable routes under changing conditions. Although this introduces additional overhead, it yields a better trade-off between latency and robustness. For example, in the 10-node setting, LeDRL improves task success rate by 12.87\% over MAPPO-TO.

	\section{Experiments in Real-World Scenarios} \label{sec:evaluation-real-world}
	
	\subsection{System Implementation and Algorithm Deployment}
	
\begin{figure}
		\begin{subfigure}{0.49\linewidth}
			\centering
			\includegraphics[clip, width=\linewidth]{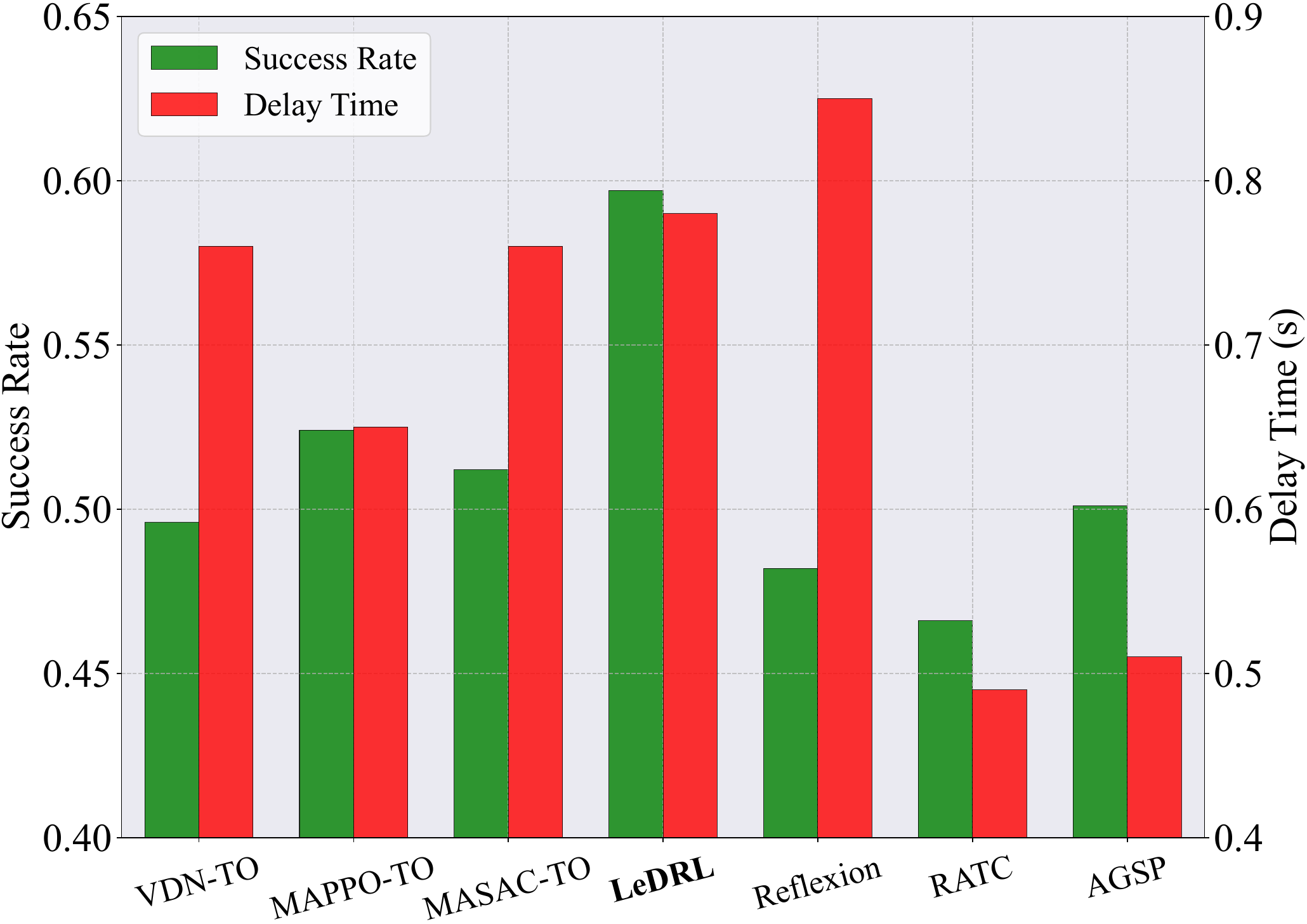}  
			\captionsetup{font={scriptsize,stretch=1.25}}
			\caption{Confidence threshold $\lambda_{\text{suc}}=0.4$}
			\label{fig:testbed1}
		\end{subfigure}
		\hfill
		\begin{subfigure}{0.49\linewidth}
			\centering
			\includegraphics[clip, width=\linewidth]{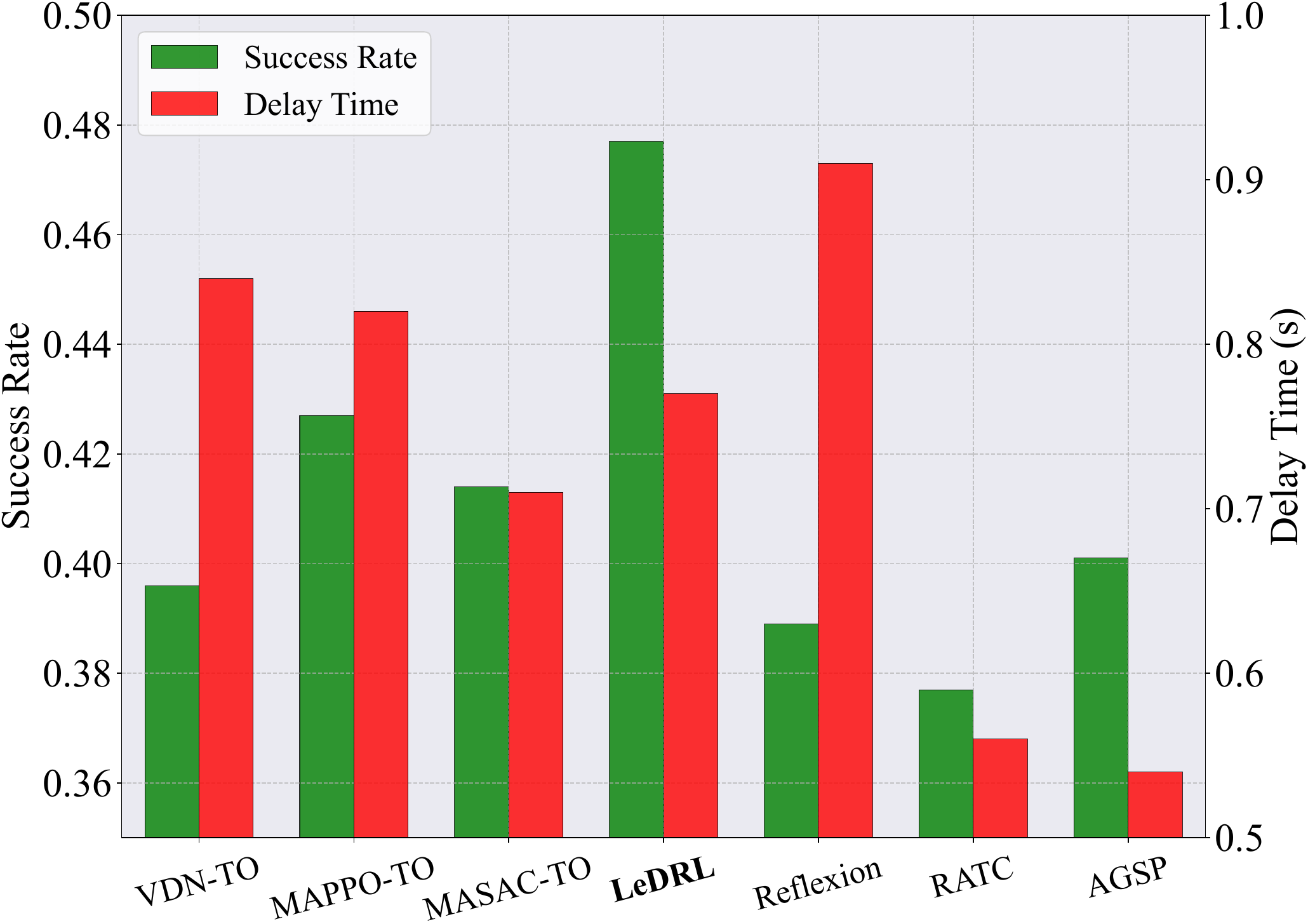}
			\captionsetup{font={scriptsize,stretch=1.25}}
			\caption{Confidence threshold $\lambda_{\text{suc}}=0.5$}
			\label{fig:testbed2}
		\end{subfigure}
		\vspace{-16pt}
		
		\captionsetup{font={scriptsize,stretch=1.25}, justification=raggedright, singlelinecheck=false}
		\caption{LeDRL success rate under different YOLO confidence ($\lambda_{\text{suc}}=0.4$ / $0.5$).}
		\label{fig:testbed}
	\end{figure}

	To evaluate deployment feasibility, we build \textit{CoEdgeSys}, a collaborative edge prototype with 10 heterogeneous Jetson nodes (4 Jetson Nano and 6 Jetson Xavier NX), all connected via a wired Gigabit LAN. Each node runs a scheduler and a YOLOv8 detector, and evaluation is conducted on the public COCO dataset. The scheduler supports LeDRL and all baselines. The actor network contains three fully connected layers and is smaller than 100KB, incurring only limited overhead on Jetson devices. For LLM guidance, \textbf{Qwen3-4B} is deployed on a server with a single RTX 4090 GPU, and CoEdgeSys supports parallel request handling, with the resulting latency reported in Fig.~\ref{fig:testbed}.
	
\begin{figure} 
		\centering
		\includegraphics[clip, width=0.45\textwidth]{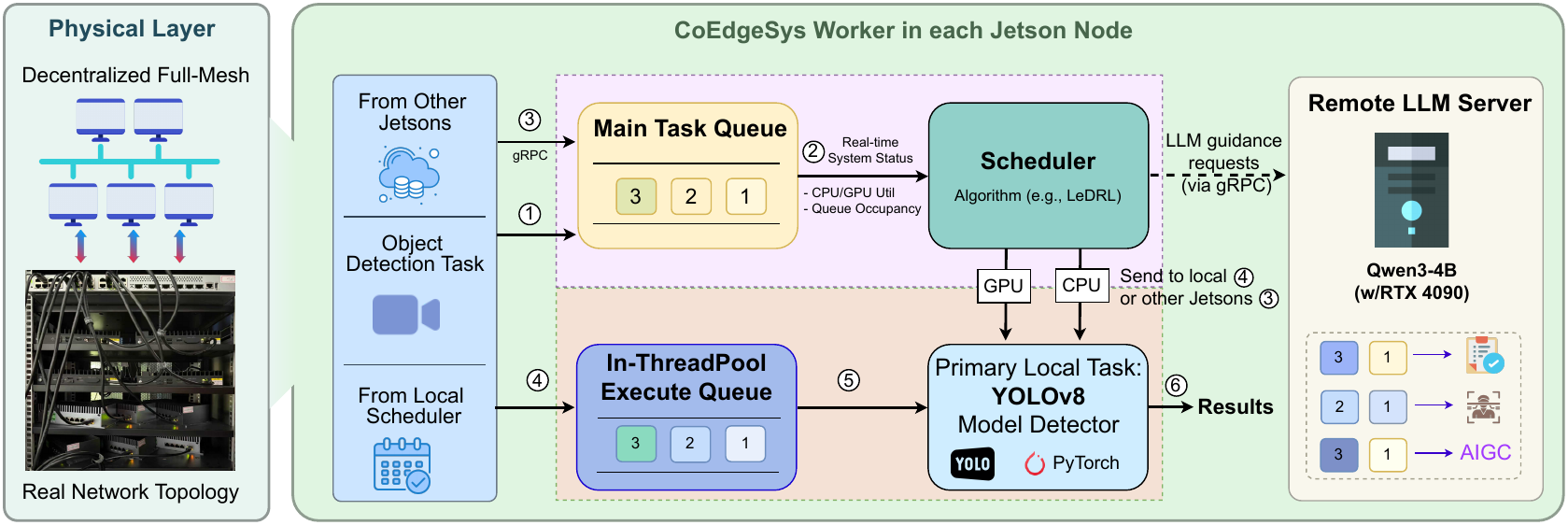}
		\captionsetup{font={scriptsize,stretch=1.25},justification=raggedright,singlelinecheck=false}
		\caption{Architecture of the CoEdgeSys prototype and its workflow on each Jetson node.}
		\label{fig:prototype_system}
		\vspace{-10pt}
	\end{figure}

	As shown in Fig.~\ref{fig:prototype_system}, at each slot $t$, tasks randomly arrive at Jetson nodes (\circlednum{1}). The local scheduler monitors real-time system status, including CPU/GPU utilization and queue occupancy, and decides whether to execute the task locally or offload it (\circlednum{2}). Local tasks enter the ThreadPool Execute Queue (\circlednum{4}), while offloaded tasks are sent via \textbf{gRPC} to a neighboring Jetson (\circlednum{3}). The assigned node runs YOLOv8, and the inference result is returned to the originating task generator (\circlednum{5} \circlednum{6}). The end-to-end delay is measured from task dispatch to result reception.
	
	\subsection{Testbed Results}
	
	To emulate real-world runtime disruptions, we randomly activate denial-of-service (DoS) states on a subset of edge nodes during specific intervals, and tasks assigned to these nodes are treated as failed. We further vary the YOLO confidence threshold to impose different application-level success criteria, where a higher threshold represents a more stringent interference setting. A strict 4-second latency constraint is also enforced, and any task exceeding it is marked as unsuccessful. We evaluate two interference settings with confidence thresholds of 0.4 (Fig.~\ref{fig:testbed1}) and 0.5 (Fig.~\ref{fig:testbed2}). In both cases, LeDRL achieves the highest task success rate (up to 60\%), outperforming \textbf{the best baseline by 13.5\%}.
	
	This advantage comes from LeDRL’s hybrid training strategy: LLM-guided policy shaping and attention-based fusion help the agent internalize failure-aware heuristics and topological sensitivity under dynamic failures, while memory and past experiences support more informed decisions than DRL-only policies. Importantly, this improvement is achieved without a significant increase in task latency, as LeDRL keeps latency close to DRL-only baselines while achieving higher success rate and better robustness under runtime disruptions.
	
	Under more stringent interference settings, baselines such as VDN-TO and RATC degrade more sharply due to limited adaptability. In contrast, LeDRL maintains stable performance, showing stronger resilience to topological shifts and runtime node failures. These results confirm that the decision-quality advantage observed in simulation transfers effectively to real-world, delay-sensitive edge environments.
	
	\section{CONCLUSION} 
	
	We present \textbf{LeDRL}, a hybrid decision framework that integrates LLM with self-attention-based DRL for decentralized task offloading in dynamic edge environments. LeDRL introduces two key innovations: a \textit{Reflective Evaluator} that converts episodic outcomes into structured feedback and retrieves relevant memory to refine decisions, and a \textit{Self-Attention Fusion} module that aligns LLM priors with learned policies to speed up convergence and improve robustness under topology and failure dynamics. Extensive simulations and a Jetson-based prototype show that LeDRL achieves superior performance over state-of-the-art baselines across multiple dimensions including success rate, adaptability, and inference efficiency.
	In future work, we will explore decoupling the LLM from online execution while preserving the robustness gains achieved during training, and we will also study scaling LeDRL to larger networks.

	\section*{Acknowledgment}
	
	We sincerely thank the anonymous reviewers for their valuable feedbacks.
	This work was supported in part by the Guangdong Basic and Applied Basic Research Foundation under Grant 2026A1515010224 and Grant 2025A1515011996, in part by Hong Kong RGC Theme-based Research Scheme (TRS) under Grant T43- 513/23-N, in part by the NSFC and Hong Kong RGC Collaborative Research Scheme under Grant 62321166652.
	
	\bibliographystyle{unsrt}
	\bibliography{references.bib}
	
\end{document}